\documentclass[prd,superscriptaddress,unsortedaddress,twocolumn,showpacs,preprintnumbers,amsmath,amssymb]{revtex4}
\usepackage{graphicx}
\usepackage{bm}
\usepackage[breaklinks, colorlinks=true, pdfstartview=FitV,linkcolor=red, citecolor=blue, urlcolor=blue]{hyperref}

\def\lessa{\mathrel{\mathpalette\fun <}}
\def\ga{\mathrel{\mathpalette\fun >}}
\def\fun#1#2{\lower3.6pt\vbox{\baselineskip0pt\lineskip.9pt
\ialign{$\mathsurround=0pt#1\hfil##\hfil$\crcr#2\crcr\sim\crcr}}}

\newcommand{\beq}{\begin{equation}}
\newcommand{\eeq}{\end{equation}}
\newcommand{\bea}{\begin{eqnarray}}
\newcommand{\eea}{\end{eqnarray}}

\begin{document}

\preprint{RBRC-966}
\title{Two-color QCD at imaginary chemical potential \\ and its impact on real chemical potential}

\author{Kouji Kashiwa}
\affiliation{RIKEN/BNL Research Center, Brookhaven National Laboratory, Upton, NY-11973, USA}

\author{Takahiro Sasaki}
\affiliation{Department of Physics, Graduate School of Sciences, Kyushu University, Fukuoka 812-8581, Japan}

\author{Hiroaki Kouno}
\affiliation{Department of Physics, Saga University, Saga 840-8502, Japan}

\author{Masanobu Yahiro}
\affiliation{Department of Physics, Graduate School of Sciences, Kyushu University, Fukuoka 812-8581, Japan}
\date{\today}

\begin{abstract}
We study properties of two-color QCD at imaginary chemical potential
($\mu$) from the viewpoint of the Roberge-Weiss (RW) periodicity, the
charge conjugation and the pseudo-reality.
At $\mu=\pm i\pi T/2$, where $T$ is temperature, the system is symmetric
under the combination of the charge conjugation ${\cal C}$ and the
${\mathbb Z}_{2}$ transformation.
The symmetry, called ${\cal C} {\mathbb Z}_{2}$ symmetry, is preserved
at lower $T$ but spontaneously broken at
higher $T$.
The Polyakov-loop extended Nambu--Jona-Lasinio (PNJL) model has the same
properties as two-color QCD for ${\cal C} {\mathbb Z}_{2}$ symmetry
and the pseudo-reality.
The nontrivial correlation between the chiral restoration and the
deconfinement are investigated by introducing the entanglement vertex in
the PNJL model.
The order of ${\cal C} {\mathbb Z}_{2}$ symmetry breaking at the RW
endpoint is second-order when the correlation is weak, but becomes
first-order when the correlation is strong.
We also investigate the impact of the correlation on the phase diagram
at real $\mu$.
\end{abstract}

\maketitle


\section{Introduction}
\label{Introduction}

Elucidation of QCD at finite temperature ($T$) and finite quark-number
chemical potential ($\mu$) is one of the most important subjects in
hadron physics.
Lattice QCD (LQCD) is the first-principle calculation, but has the sign
problem at real $\mu$.
Particularly at $\mu/T \gtrsim 1$, the LQCD calculation is not feasible,
although several methods have been proposed so far to circumvent the
problem; see for example Ref.~\cite{Forcrand:2009}.
For this reason, effective models such as the Polyakov-loop extended
Nambu--Jona-Lasinio (PNJL)
model~\cite{Fukushima:2004,Ratti:2006,Sakai:2008,Hell:2009,Kashiwa:2011}
are widely used to investigate QCD at finite $\mu$.

Symmetries are important to understand QCD.
In this paper we focus our discussion on global symmetries.
QCD has chiral symmetry in the limit of zero current quark mass ($m$)
and $\mathbb{Z}_{N_c}$ symmetry in the limit of infinite $m$, where
${N_c}$ is the number of colors.
Charge conjugation (${\cal C}$) symmetry is preserved at $\mu=0$ but not
at finite $\mu$, since the QCD action $S(\mu/T,T)$ is transformed by
${\cal C}$ as
\bea
S(\mu/T,T) \xrightarrow{\cal C} S(-\mu/T,T).
\label{C-transformation}
\eea
This indicates that
\begin{align}
Z(\mu/T,T)=Z^{(c)}(\mu/T,T)=Z(-\mu/T,T),
\label{CC-relation}
\end{align}
where $Z^{(c)}$ is the partition function written with the ${\cal
C}$-transformed quark and gauge fields and the second equality comes
from \eqref{C-transformation}.
Thus ${\cal C}$ symmetry is broken at finite $\mu$, but it derives the
fact that $Z(\mu/T,T)$ is $\mu$-even.

Similar discussion is possible also for imaginary chemical potential
$\mu=i \theta T$, where $\theta$ is the dimensionless imaginary chemical
potential.
For simplicity, we use $S(\theta)$ and $Z(\theta)$ as shorthand
notations of $S(i\theta,T)$ and $Z(i\theta,T)$, respectively.
Consider the transformation
\begin{eqnarray}
q(x,\tau) \to \exp{(i\theta T\tau )}q(x,\tau)
\label{transform_1}
\end{eqnarray}
for the quark field $q$, 
where ${\bf x}$ and $\tau$ are the spatial and imaginary time variables.
After this transformation, the action $S(\theta)$ with the standard
boundary condition $q(x, \beta)=-q(x, 0)$ is changed into $S(0)$ with
the twisted boundary condition (TBC)
\bea
q(x, \beta)=-R(\theta)q(x, 0)
\label{TBC}
\eea
with the twist factor
\bea
R(\theta)=\exp{(-i \theta)}.
\label{R-factor}
\eea
The action $S(0)$ with the twist factor \eqref{R-factor} in its quark
boundary condition is a good starting point to understand roles of
${\cal C}$ and $\mathbb{Z}_{N_c}$ at finite $\theta$.
For example, one can easily see from \eqref{R-factor} that $Z(\theta)$
has a periodicity of $2\pi$.
We can then consider one period of either $0 \le \theta < 2\pi$ or $-\pi
\le \theta < \pi$.

The action $S(0)$ is not transformed by ${\cal C}$, but the twist factor
is changed as
\bea
R(\theta) \xrightarrow{\cal C} R(-\theta) .
\label{TBC-C}
\eea
This indicates that
\bea
Z(\theta)=Z^{(c)}(\theta)=Z(-\theta).
\label{CC-relation-Im}
\eea

The action $S(0)$ is also invariant under the $\mathbb{Z}_{N_c}$
transformation
\begin{align}
q \to U q, ~
A_{\nu} \to UA_{\nu}U^{-1} -{i} (\partial_{\nu}U)U^{-1},
\label{z3-trans}
\end{align}
where $A_\nu$ is the gauge field and $U(x,\tau)$ are elements of
SU($N_c$) with the boundary condition
$
U(x,\beta=1/T)=\exp(-2i \pi k/N_c)U(x,0)
$
for integers $k=0, \cdots, N_\mathrm{c}-1$.
However the $\mathbb{Z}_{N_c}$ transformation
changes $R(\theta)$ as~\cite{Roberge:1986}
\bea
R(\theta) \xrightarrow{\mathbb{Z}_{N_c}} R(\theta- 2\pi k/N_c) .
\label{TBC-Z}
\eea
This indicates that
\begin{align}
Z(\theta)=Z^{(z)}(\theta)=Z(\theta-2{\pi}k/N_c) ,
\label{RW-periodicity_original}
\end{align}
or equivalently
\begin{align}
Z(\theta)=Z(\theta+2{\pi}k/N_c) ,
\label{RW-periodicity}
\end{align}
where $Z^{(z)}$ is the partition function written with the ${\mathbb
Z}_{N_c}$-transformed quark and gauge fields.
Either \eqref{RW-periodicity_original} or \eqref{RW-periodicity} is
called the Roberge-Weiss (RW) periodicity~\cite{Roberge:1986}.
The periodicity means that $Z(\theta)$ is invariant under the
combination of the ${\mathbb Z}_{N_c}$ transformation and the parameter
transformation $\theta \to \theta + 2 \pi k/N_c$, i.e., under the
extended ${\mathbb Z}_{N_c}$ transformation \cite{Sakai:2008}.

At imaginary chemical potential, the ${\cal C}$ and ${\mathbb Z}_{N_c}$
thus break down through the quark boundary condition.
As shown below, however, $R(\theta)$ is invariant under the ${\cal C}$
transformation or the combination of the ${\cal C}$ and
${\mathbb Z}_{N_c}$ transformations at special values of $\theta$.
At $\theta=\pi$, $R(\theta)$ is ${\cal C}$-invariant, since
$R(\pi)=R(-\pi)$.
At $\theta=\pi/N_\mathrm{c}$, $R(\theta)$ is invariant under the
combination of the ${\cal C}$ transformation and the ${\mathbb Z}_{N_c}$
transformation with $k=-1$:
\bea
R(\pi/N_\mathrm{c}) \xrightarrow{\cal C}
R(-\pi/N_\mathrm{c}) \xrightarrow{{\mathbb Z}_{N_c}}
R(\pi/N_\mathrm{c}) .
\label{CZN-transfomation}
\eea
QCD has the same symmetry at $\theta=\pi/N_\mathrm{c}$ mod
$2\pi/N_\mathrm{c}$ because of the RW periodicity. We refer to this
symmetry as ${\cal C}{\mathbb Z}_{N_c}$ symmetry in this paper,
particularly when the ${\mathbb Z}_{N_c}$ transformation used is not the
identity transformation.
An order parameter of the symmetry is a
${\cal C}$-odd and ${\mathbb Z}_{N_c}$-invariant quantity such as the
quark number density $n_q$.
In this sense ${\cal C}{\mathbb Z}_{N_c}$ symmetry has properties
similar to
${\cal C}$ symmetry.

For $N_\mathrm{c}=3$ as a typical case of odd $N_\mathrm{c}$, there
appears ${\cal C}{\mathbb Z}_{3}$ symmetry at
$\theta=\pm \pi/3, \pi$. Particularly at $\theta=\pi$, the symmetry is
reduced to ${\cal C}$ symmetry.
${\cal C}{\mathbb Z}_{3}$ symmetries at
$\theta=\pm \pi/3$ can be understood as ${\mathbb Z}_{3}$ images of
${\cal C}$ symmetry at $\theta=\pi$.
The symmetries at $\theta=0, \pm \pi/3, \pi$ are summarized in Table
\ref{Table:sym3}.
Note that these symmetries may be spontaneously broken in some cases
as shown below.
Also for $N_\mathrm{c}=2$ as a typical case of even $N_\mathrm{c}$, the
system has ${\cal C}{\mathbb Z}_{2}$ symmetry at
$\theta=\pm \pi/2$ and ${\cal C}$ symmetry at $\theta=0, \pi$; see Table
\ref{Table:sym} for the summary of symmetries.
${\cal C}{\mathbb Z}_{2}$ symmetries at $\theta= \pm \pi/2$ are,
however, not ${\mathbb Z}_{2}$ images of ${\cal C}$ symmetries at
$\theta=0, \pi$
because the number of ${\mathbb Z}_{2}$ images are $2$.
Further understanding is thus necessary for ${\cal C}{\mathbb
Z}_{N_\mathrm{c}}$ symmetry with even $N_\mathrm{c}$.

\begin{table}[h]
\begin{center}
\begin{tabular}{cccc}
\\
\hline
\hline
~~$\theta$~~ & ~~${\mathbb Z}_{3}$~~ & $C$ & ${\cal C}{\mathbb Z}_{3}$ \\
\hline
\hline
0 &~\phantom{Invariant}~&~Invariant~&~\phantom{Invariant}~ \\
\hline
~~$\pm \pi/3$~~ & & & Invariant \\
\hline
$\pi$ & & Invariant & \\
\hline
\hline
\end{tabular}
\end{center}
\caption{
Invariances of twist factor $R(\theta)$ in three-color QCD. 
${\cal C}{\mathbb Z}_3$ transformation is defined with $k=\mp 1$ for $\theta =\pm \pi/3$.
Note that $S(0)$ is invariant under both $\mathbb{Z}_3$ and
$C$ transformations.
}
\label{Table:sym3}
\end{table}

\begin{table}[h]
\begin{center}
\begin{tabular}{cccc}
\\
\hline
\hline
~~$\theta$~~ & ~~${\mathbb Z}_{2}$~~ & $C$ & ${\cal C}{\mathbb Z}_{2}$ \\
\hline
\hline
0, $\pi$ &~\phantom{Invariant}~&~Invariant~&~\phantom{Invariant}~ \\
\hline
~~$\pm \pi/2$~~ &  & & Invariant \\
\hline
\hline
\end{tabular}
\end{center}
\caption{
Invariances of twist factor $R(\theta)$ in two-color QCD.
${\cal C}{\mathbb Z}_2$ transformation is defined with $k=\mp 1$ for $\theta =\pm \pi /2$.
Note that $S(0)$ is invariant under both $\mathbb{Z}_2$ and $C$ transformations.
}
\label{Table:sym}
\end{table}

When $T$ is higher than some temperature $T^c_{\rm RW}$, there appears a
first-order phase transition at
$\theta=\pi/N_\mathrm{c}$ mod $2\pi/N_\mathrm{c}$~\cite{Roberge:1986}.
Just on the transition line, the spontaneous breaking of either
${\cal C}$ or ${\cal C}{\mathbb Z}_{N_c}$ symmetry takes
place~\cite{Kouno:2009}.
The transition is now called the RW phase transition.
A current topic on the RW phase transition is the order of the
transition at its endpoint, i.e., the RW endpoint.
Recent three-color LQCD simulations show that the order is first-order
for small and large $m$, but second-order for intermediate
$m$~\cite{D'Elia:2009,Bonati:2011a,Forcrand:2010,Bonati:2011b}.
The order may be first-order for both
two-flavor~\cite{D'Elia:2009,Bonati:2011a} and three-flavor
cases~\cite{Forcrand:2010,Bonati:2011b}, when the pion mass $m_{\pi}$
has the physical value.
If the order is first-order, the RW endpoint becomes a triple-point at
which three first-order transition lines meet.
The PNJL model reproduces these
results~\cite{Sakai:2010qc,Sakai:2010,Sasaki:2011}.

Two-color QCD has some interesting points.
The number of colors, $N_c$, can vary from 2 to infinity, that is,
realistic three-color QCD is between two-color QCD and large $N_c$ QCD.
In this sense, understanding of both two-color and large-$N_c$ QCD is
important.
The algebraic approach based on the pseudo-reality~\cite{Kogut:2000}
plays an important role in two-color QCD, while large-$N_c$ QCD is well
understood by the geometric approach based on the $1/N_c$
expansion~\cite{Witten:1979} or the AdS/CFT
correspondence~\cite{Maldacena:1997re}.
In virtue of the pseudo-reality, two-color LQCD has no sign problem not
only at imaginary $\mu$ but also at real $\mu$~\cite{Kogut:2001}, and
consequently LQCD data are available
there~\cite{Giudice:2004,Cea:2007,Cea:2007ks,Cea:2008,Cea:2009ud,Cea:2009}.
Furthermore, two-color QCD has higher symmetry at imaginary $\mu$ than
at real $\mu$, that is
${\cal C}{\mathbb Z}_{2}$ symmetry at $\theta=\pm \pi/2$.

In this paper, we study properties of two-color QCD at imaginary $\mu$
from the viewpoint of the RW periodicity, ${\cal C}{\mathbb Z}_{2}$
symmetry and the pseudo-reality.
The PNJL model has the same properties as two-color QCD for the RW
periodicity, ${\cal C}{\mathbb Z}_{2}$ symmetry and the pseudo-reality.
The PNJL model is then used to investigate two-color QCD concretely.
Particularly, the nontrivial correlation between chiral and ${\cal
C}{\mathbb Z}_{2}$ symmetry breakings are investigated.

This paper is organized as follows.
In Sec. \ref{Two-color QCD}, some properties of two-color QCD are
derived with the ${\mathbb Z}_{2}$ transformation, the charge
conjugation and the pseudo-reality.
In Sec. \ref{Formalism}, the two-color PNJL model is formulated with the
mean-field approximation.
Numerical results are shown in Sec. \ref{Numerical results}.
Section \ref{Summary} is devoted to summary.

\section{Properties of two-color QCD}
\label{Two-color QCD}

We first consider the one-flavor ($N_f=1$) case.
The partition function $Z$ of two-color QCD is obtained in Euclidean
spacetime as
\begin{align}
Z=\int DA \det[M(\mu)] \exp[-{1\over{4g^2}}F_{\mu\nu}^2]
\label{QCD-Z}
\end{align}
with
\begin{align}
M(\mu)=D + m - \gamma_4 \mu,
\end{align}
where the Dirac operator $D$ is defined by
$D=\gamma_{\mu}(\partial_{\mu}-iA_{\mu})$ for the current quark mass $m$
and the gauge field $A_\nu$, and
$F_{\mu\nu}=\partial_\mu A_\nu -\partial_\nu A_\mu -i[A_\mu ,A_\nu ]$.
It is assumed that $\mu$ is either real or pure imaginary.
For later convenience, we define Pauli matrices $t_i$ in color space and
the Dirac charge-conjugation matrix $C=\gamma_2\gamma_4$.

The fermion determinant $\det[M(\mu)]$ satisfies
\begin{align}
(\det[M(\mu)])^*=\det[M(-\mu^*)],
\label{M-relation-1}
\end{align}
since
\begin{align}
(\det[M(\mu)])^*=\det[M(\mu)^{\dagger}]
&=\det[\gamma_5M(\mu)^{\dagger}\gamma_5]
\notag \\
&=\det[M(-\mu^*)] .
\end{align}
The relation \eqref{M-relation-1} indicates that $\det[M(\mu)]$ is real
when $\mu$ is pure imaginary.
The relation \eqref{M-relation-1} is true for any $N_c$.

Two-color QCD has the pseudo-reality~\cite{Kogut:2000},
\begin{align}
D t_2 C \gamma_5= t_2 C \gamma_5 D^* ,
\label{eq:pseudo-reality}
\end{align}
and hence the fermion determinant satisfies
\begin{align}
\det[M(\mu)]&=\det[(t_2 C \gamma_5)^{-1}M(\mu)(t_2 C \gamma_5)]
\notag \\
&=(\det[M(\mu^*)])^*
\label{eq:pseudo-reality-2}
\end{align}
in virtue of the pseudo-reality \eqref{eq:pseudo-reality}.
This means that $\det[M(\mu)]$ is real when $\mu$ is real.
Two-color LQCD thus has no sign problem at both real and pure imaginary
$\mu$~\cite{Kogut:2001}.
The charge-conjugation relation \eqref{CC-relation} is obtained from
\eqref{M-relation-1} and \eqref{eq:pseudo-reality-2}.

The Polyakov loop $\Phi$ is the vacuum expectation value
of the Polyakov-loop operator
\begin{align}
L = {1\over{N_c}}{\rm tr}_c
\Bigl({\cal P}
\exp\Bigl[i \int_0^{1/T} d \tau A_4 \Bigr]
\Bigr) .
\label{Polyakov_1}
\end{align}
with the time-ordering operator ${\cal P}$.
The operator $L$ is real in the two-color system, because
\begin{align}
L^* = {1\over{2}}{\rm tr}_c
\Bigl({\cal P}
\exp\Bigl[-i \int_0^{1/T} d\tau (A_4)^{*} \Bigr]
\Bigr)
=L ,
\label{Polyakov_2}
\end{align}
where the second equality is obtained from the identity
$ t_2 A_{\nu} t_2=-(A_{\nu})^{\rm T}=-(A_{\nu})^*$.
The Polyakov loop $\Phi$ is hence real at both real and pure imaginary
$\mu$; note that $\det[M(\mu)]$ is real there.
Under the charge conjugation, the factor $-(A_4)^{\rm T}$ is transformed
into $A_4$. Therefore the second equality of \eqref{Polyakov_2} means
that $L$ is ${\cal C}$-invariant.
Using this property and the relation \eqref{CC-relation}, one can see that
\begin{align}
\Phi(\mu/T,T)=\Phi(-\mu/T,T).
\label{Phi-even}
\end{align}

For pure imaginary chemical potential $\mu=iT\theta$, it is convenient
to introduce the modified Polyakov loop
\bea
\Psi(\theta) \equiv \Phi(\theta) e^{i\theta},
\label{Psi-def}
\eea
where $\Phi(\theta)$ has been used as a shorthand notation of
$\Phi(i \theta,T)$.
The modified Polyakov loop satisfies the RW periodicity
\bea
\Psi(\theta)=\Psi(\theta+\pi),
\label{RW-periodicity-two-color}
\eea
since it is invariant under the extended ${\mathbb Z}_{N_c}$
transformation~\cite{Sakai:2008}. Inserting \eqref{Psi-def} into
\eqref{RW-periodicity-two-color} leads to
\bea
\Phi(\theta)= - \Phi(\theta+\pi) .
\label{RW-periodicity-Phi}
\eea
One can also see from \eqref{Phi-even}, \eqref{Psi-def} and
$\Phi^*(\theta)=\Phi(\theta)$ that
\bea
\Phi(\theta)=\Phi(-\theta),~~~\Psi(\theta)^*=\Psi(-\theta) .
\label{Psi-relation}
\eea
Hence the imaginary part ${\rm Im}[\Psi(\theta)]$ is
$\theta$-odd, whereas the real part ${\rm Re}[\Psi(\theta)]$ and
$\Phi(\theta)$ are $\theta$-even.

At $\theta=\pm \pi/2$, two-color QCD has ${\cal C} {\mathbb Z}_{2}$
symmetry, as mentioned in Sec.~\ref{Introduction}.
And $\Phi(\theta)$ and ${\rm Im}[\Psi(\theta)]$ are order parameters of
the symmetry, since
$\Phi(\theta)$ and ${\rm Im}[\Psi(\theta)]$ are
${\cal C} {\mathbb Z}_{2}$-odd at $\theta=\pm \pi/2$:
\bea
&&\Phi(\theta) \xrightarrow{\cal C}
\Phi(-\theta) \xrightarrow{{\mathbb Z}_{2}}
-\Phi(\theta) ,\\
&&{\rm Im}[\Psi(\theta)] \xrightarrow{\cal C}
{\rm Im}[\Psi(-\theta)] \xrightarrow{{\mathbb Z}_{2}}
-{\rm Im}[\Psi(\theta)] ,~~~~~
\eea
because $L$ is ${\cal C}$-invariant and transformed by the
${\mathbb Z}_{2}$ transformation as
$
L \to - L.
$
The two order parameters are identical with each other at $\theta=\pm
\pi/2$, since ${\rm
Im}[\Psi(\theta)]=\Phi(\theta)\sin(\theta)=\Phi(\theta)$.
This is not surprising, because two-color QCD has only one symmetry there.
At $\theta=0, \pi$, meanwhile, QCD has ${\cal C}$ symmetry and the order
parameter is a ${\cal C}$-odd quantity such as $n_q$.
Table \ref{Table:sym} shows the symmetries that two-color QCD has at
$\theta=0, \pm \pi/2, \pi$.

Next we consider the $N_f=2$ case. The fermion determinant is described by
\begin{align}
\det[M(\mu)]=\det[M_u(\mu)]\det[M_d(\mu)],
\label{M-relation-Nf=2}
\end{align}
where $\det[M_u(\mu)]$ and $\det[M_d(\mu)]$ are the fermion determinants
for u- and d-quark, respectively. Using the operator $C t_2$ only for
$\det[M_d(\mu)]$, one can get the relation
\begin{align}
\det[M(\mu)]&=\det[M_u(\mu)]\det[(C t_2)^{-1} M_d(\mu)(C t_2)]
\notag \\
&=\det[M_u(\mu)]\det[M_d(-\mu)].
\label{M-relation-Nf=2-2}
\end{align}
This relation indicates that the $1+1$ system with finite $\mu$ is
identical with the $1+1^*$ system with the same amount of isospin
chemical potential
$\mu_{\rm iso}$. The diquark condensate in the former system corresponds
to the pion condensate in the latter system.
Because of this symmetry, we consider the former system only in the
present paper.

Two-color LQCD simulations were made in
Ref.~\cite{Giudice:2004,Cea:2007,Cea:2008,Cea:2009} for the $N_f=8$ case.
The LQCD results at $\theta=\pi/2$ show that
$\Phi=0$ at small $T$ but finite at large $T$.
This indicates that the spontaneous breaking of ${\cal C}{\mathbb
Z}_{2}$ symmetry occurs at some temperature $T^c_{\rm RW}$.
Further analyses are made in Sec. \ref{Formalism} and \ref{Numerical
results} by using the PNJL model.

\section{PNJL model}
\label{Formalism}

We consider the $N_c=N_f=2$ case.
The PNJL Lagrangian is obtained in Minkowski spacetime by
\begin{align}
{\cal L}
&= {\bar q} ( i \gamma^\nu D_\nu - m ) q
\nonumber\\
&+ G [ ({\bar q}q)^2 + ({\bar q} i\gamma_5 {\vec \tau} q)^2
+
| q^T C i\gamma_5 \tau_2 t_2 q |^2
]
\nonumber\\
&
- G_\mathrm{v} ({\bar q} \gamma^\mu q)^2
- {\cal U}(\Phi ),
\label{PNJL-L}
\end{align}
where $q$ is the two-flavor quark field, $m$ is the current quark mass and
$t_i$ and $\tau_i$ are Pauli matrices in color and flavor spaces,
respectively.
In the limit of $m=\mu=0$, two-color QCD has Pauli-G\"ursey
symmetry~\cite{Pauli:1957,Guersey:1958}, so the PNJL Lagrangian is so
constructed as to have the symmetry.
Note that the vector-type four-quark interaction $({\bar q} \gamma^\mu
q)^2$ does not work at $\mu=0$ in the mean-field level, since the
vector-type condensate is zero at $\mu=0$.
The potential ${\cal U}$ is a function of the Polyakov loop $\Phi$ and
the explicit form is shown later in Sec. \ref{Numerical results}.

Using the mean-field approximation, one can get the effective potential $\Omega$ as \cite{Brauner:2009}
\begin{align}
\Omega
&= - 2 N_\mathrm{f} \int \frac{d^3 p}{(2\pi)^3}
\sum_{\pm}
\Bigl[
\frac{1}{2} N_\mathrm{c} E_{p}^{\pm} + T ( \ln f^- + \ln f^+)
\Bigr]
\nonumber\\
& + U + {\cal U}(\Phi )
\label{TP-PNJL}
\end{align}
with
\begin{align}
f^\pm &= 1 + 2 \Phi e^{-\beta E^\pm_{p}}
+ e^{-2\beta E^\pm_{p}} ,
\\
U &= G( \sigma^2 + \Delta^2) - G_\mathrm{v} n_q^2
\end{align}
for the chiral condensate $\sigma=\langle {\bar q} q \rangle$, the
diquark condensate
$\Delta=|\langle q^T C i\gamma_5 \tau_2 t_2 q \rangle|$ and the vector
condensate (quark number density) $n_q = \langle q^\dag q \rangle$.
Here the factors $E^\pm_{p}$ are defined by
\begin{align}
E^\pm_{p} &= sgn(E_{p} \pm \mu_\mathrm{v}) \sqrt { ( E_{p} \pm
\mu_\mathrm{v})^2 +
( 2 G \Delta)^2 },
\end{align}
for finite $\Delta$ and
\begin{align}
E^\pm_{p} = E_{p} \pm \mu_\mathrm{v}
\end{align}
for $\Delta=0$, where $sgn(E_{p} \pm \mu_\mathrm{v})$ is the sign function,
$E_{p} = \sqrt{p^2 + M^2}$,
$M = m - 2 G \sigma$ and
$\mu_\mathrm{v} = \mu - 2 G_\mathrm{v} n_q$.
In the limit of $m=\mu=0$, the condensate $n_q$ is zero, so that
$\Omega$ becomes invariant under the rotation in the $\sigma$-$\Delta$
plane as a consequence of Pauli-G\"ursey symmetry.

In the Polyakov gauge, the Polyakov-loop $\Phi$ is obtained by
\begin{align}
\Phi &= \frac{1}{2} \Bigl( e^{i \phi} + e^{-i \phi} \Bigl) = \cos(\phi)
\end{align}
for real number $\phi$, indicating that $\Phi$ is real.
The mean fields $X=\sigma, \Delta, n_q, \Phi$ are determined from the
stationary conditions
\bea
\frac{\partial \Omega}{\partial X}=0,
\label{stationary-condition}
\eea
where $\Omega$ is regularized by the three-dimensional momentum cutoff
\begin{equation}
\int \frac{d^3 p}{(2 \pi)^3}\to
{1\over{2\pi^2}} \int_0^\Lambda dp p^2,
\end{equation}
because this model is nonrenormalizable.

At imaginary chemical potential $\mu=i\theta T$,
$\Omega$ is invariant under the extended ${\mathbb Z}_{2}$
transformation \cite{Sakai:2008}
\bea
\Phi \to e^{-i \pi} \Phi, \quad \theta \to \theta + \pi.
\eea
This can be understood easily by introducing the modified Polyakov-loop
$\Psi = e^{i\theta} \Phi$ and its conjugate $\Psi^* = e^{-i\theta} \Phi$
invariant under the extended ${\mathbb Z}_{2}$ transformation.
The condensate $\Delta$ is zero at imaginary $\mu$, since
$\Delta$ becomes finite only for $\mu^2 \ge
m_{\pi}^2/4$~\cite{Kogut:2000,Sakai:2010qc}.
When $\Delta=0$, $\Omega$ is rewritten into the form of \eqref{TP-PNJL}
with
\begin{align}
f^+ &= 1 + 2\Psi e^{-2i \theta} e^{-\beta (E_{p} - 2 G_\mathrm{v} n_q)}
+ e^{-2\beta E^+_{p}} ,
\label{f+}
\\
f^- &= 1 + 2\Psi^* e^{2i \theta} e^{-\beta (E_{p} + 2 G_\mathrm{v} n_q)}
+ e^{-2\beta E^-_{p}} .
\label{f-}
\end{align}
Obviously, Eqs. \eqref{f+} and \eqref{f-} show that
$\Omega$ is invariant under the extended ${\mathbb Z}_{2}$
transformation and at the same time that $\Omega$ has a periodicity of
$\pi$ in $\theta$,
i.e., the RW periodicity
\bea
\Omega(\theta)=\Omega(\theta+\pi) .
\eea
It should be noted that if $\Delta$ is finite, $\Omega$ will not have
the RW periodicity.
Since $\theta$-dependence of mean fields
$X=\sigma, n_q, \Psi$ are determined from $\Omega$ by the stationary
conditions \eqref{stationary-condition}, all the $X$ have the RW periodicity
\bea
X(\theta)=X(\theta+\pi) .
\label{X-theta}
\eea
Furthermore, the RW periodicity $\Psi(\theta)=\Psi(\theta+\pi)$
yields the relation
\bea
\Phi(\theta)=-\Phi(\theta+\pi).
\label{Phi-theta}
\eea

The PNJL Lagrangian ${\cal L}$ is invariant under the combination of the
${\cal C}$ and the parameter transformation $\theta \to -\theta$.
This property guarantees that $\Omega(\theta)=\Omega(-\theta)$ and
thereby ${\cal C}$-even quantities $\sigma$, $\Delta$ and $\Phi$ are
$\theta$-even, whereas ${\cal C}$-odd quantity $n_q$ is $\theta$-odd.
In the PNJL Lagrangian, $\theta$ appears only through the pure imaginary
factor $i\theta$. This shows that
$\theta$-odd quantities become pure imaginary, while $\theta$-even
quantities become real.
The second equation of \eqref{Psi-relation} is obtainable from the fact
that
$\Phi(\theta)^*=\Phi(\theta)=\Phi(-\theta)$, namely
\bea
\Psi(\theta)^*=(\Phi(\theta)e^{i\theta})^*=\Phi(\theta)^* e^{-i\theta}
=\Psi(-\theta).
\label{Psi-relation-2}
\eea
The PNJL model thus has the same properties as two-color QCD for the RW
periodicity and the $\theta$-parity.

Using \eqref{Phi-theta} and $\Phi(\theta)=\Phi(-\theta)$, one can see that
\bea
\Phi(\theta)=-\Phi(\pi-\theta)
\label{Phi-Z2}
\eea
and hence $\Phi(\theta)=0$ at $\theta=\pi/2$,
if $\Phi$ is a smooth function of $\theta$.
When $T$ is larger than some temperature $T_{\rm RW}$, $\Omega$ is not a
smooth function of $\theta$ at $\theta=\pi/2$.
This is the RW phase transition.
Once the RW phase transition occurs, $\Phi(\theta)$ is not smooth at
$\theta=\pi/2$ and consequently $\Phi(\theta)$ becomes finite there.
This means that the spontaneous breaking of ${\cal C}{\mathbb Z}_{2}$
symmetry takes place on the RW phase transition line.

The relations \eqref{X-theta} and \eqref{Psi-relation-2}, meanwhile, yield
a similar relation to \eqref{Phi-Z2}:
\bea
{\rm Im}[\Psi(\theta)]=-{\rm Im}[\Psi(\pi-\theta)]
\label{Im-Phi-Z2}
\eea
At $\theta=\pm \pi/2$, therefore, ${\rm Im}[\Psi]$ serves as an order
parameter of ${\cal C}{\mathbb Z}_{2}$ symmetry as well as $\Phi$.

\section{Numerical results}
\label{Numerical results}

\subsection{Parameter setting}
\label{Parameter setting}

For the $N_c=N_f=2$ case, we do not have enough LQCD data at imaginary
chemical potential. We then make qualitative analyses here.
It is well known from the nonlocal version of the PNJL model
~\cite{Blaschke:2008ui,Contrera:2008hs,Kondo:2010,Hell:2009,Hell:2011,Kashiwa:2011,Pagura:2011}
that there is the correlation between the chiral order-parameter and the
Polyakov-loop in the coupling constant through the distribution function.
This feature can be phenomenologically introduced in the local PNJL
model by using the entanglement vertex~\cite{Sakai:2010}.
This entanglement is taken into account in the present analysis.
We make the following parameter setting.

\begin{enumerate}
\item
Since the ratio $r \equiv G_\mathrm{v}/G$ is of order $(N_c)^0$ in the
leading order of the $1/N_c$ expansion, we take $r=0.4$ that is
determined from the $N_c=3$ case by comparing the result of the nonlocal
PNJL model with LQCD data at finite imaginary chemical potential.
\item
We introduce the entanglement vertex of the form $G_i (1 - \alpha
\Phi^2)$ for $G_i=G$ and $G_\mathrm{v}$, respecting ${\mathbb
Z}_{2}$ symmetry.
Here the entanglement parameter $\alpha$ is treated as a free parameter,
but is assumed to be common for both $G$ and $G_\mathrm{v}$.
Vacuum properties are unchanged for any value of $\alpha$.
We mainly use $\alpha=0.4$ at which pseudo-critical temperatures of the
chiral and deconfinement transitions almost coincide when $\mu=0$.
\item
In the leading order of the $1/N_c$ expansion,
$m_{\pi}$ is scaled by $(N_c)^0$ and the pion decay constant $f_{\pi}$
by $\sqrt{N_c}$. These scaling properties are assumed to determine
the parameter set $(G, \Lambda, m)$ of the NJL sector, where $m$ is
simply assumed to be 5.4~MeV. In this parameter set, the dynamical quark
mass $M$ becomes $M_0=305$~MeV at $T=\mu=0$.
The resulting parameters are shown in Table~\ref{Table}, together with
the values of $m_{\pi}$, $f_{\pi}$ and $M_0$.
\item
Following Ref.~\cite{Brauner:2009}, we take the Polyakov-loop effective
potential of the form
\begin{align}
\frac{{\cal U} (\Phi) }{T}
&= -b \Bigl[ 24 e^{-a/T} \Phi^2 + \ln (1-\Phi^2) \Bigr]
\end{align}
with $a = 858.1$ MeV and $b^{1/3} = 210.5$ MeV.
This potential yields a second-order deconfinement transition in the
pure gauge limit.
\end{enumerate}

In Ref.~\cite{Morita:2011eu}, the scalar coupling $G$ was varied to
investigate the effect on the phase transition. In this case, however,
the change of $G$ also varies
vacuum properties such as $m_\pi$ and $f_\pi$.
We therefore fix $G$ in this paper not to change the vacuum properties.
The Polyakov-loop effective potential ${\cal U}$ used here is determined
by using the strong coupling expansion of the pure Yang-Mills theory.
The logarithmic part comes from the Haar measure.
Since the parameter fitting procedure does not refer to microscopic
dynamics, it is unclear how the Polyakov-loop effective potential is
related to non-perturbative characteristics near $T_c$.
This problem should be investigated elsewhere by considering other
approach based on gluon and ghost propagators~\cite{Fukushima:2012}.

\begin{table}[h]
\begin{center}
\begin{tabular}{ccc}
\\
\hline
\hline
$ m_\pi$ [MeV]~~ & $ f_\pi$ [MeV]~~ & $M_0$ [MeV]~~ \\
\hline
140 & 75.4 & 305 \\
\hline
\hline
$ G $ [GeV$^{-2}$] & $\Lambda$ [MeV] & $m $ [MeV]\\
\hline
7.23 & 657 & 5.4\\
\hline
\hline
\end{tabular}
\caption{ Summary of parameters and physical values.
}
\label{Table}
\end{center}
\end{table}

\subsection{$\theta$ dependence of order parameters}
\label{Transitions}

\begin{figure}[htbp]
\begin{center}
\includegraphics[width=0.23\textwidth]{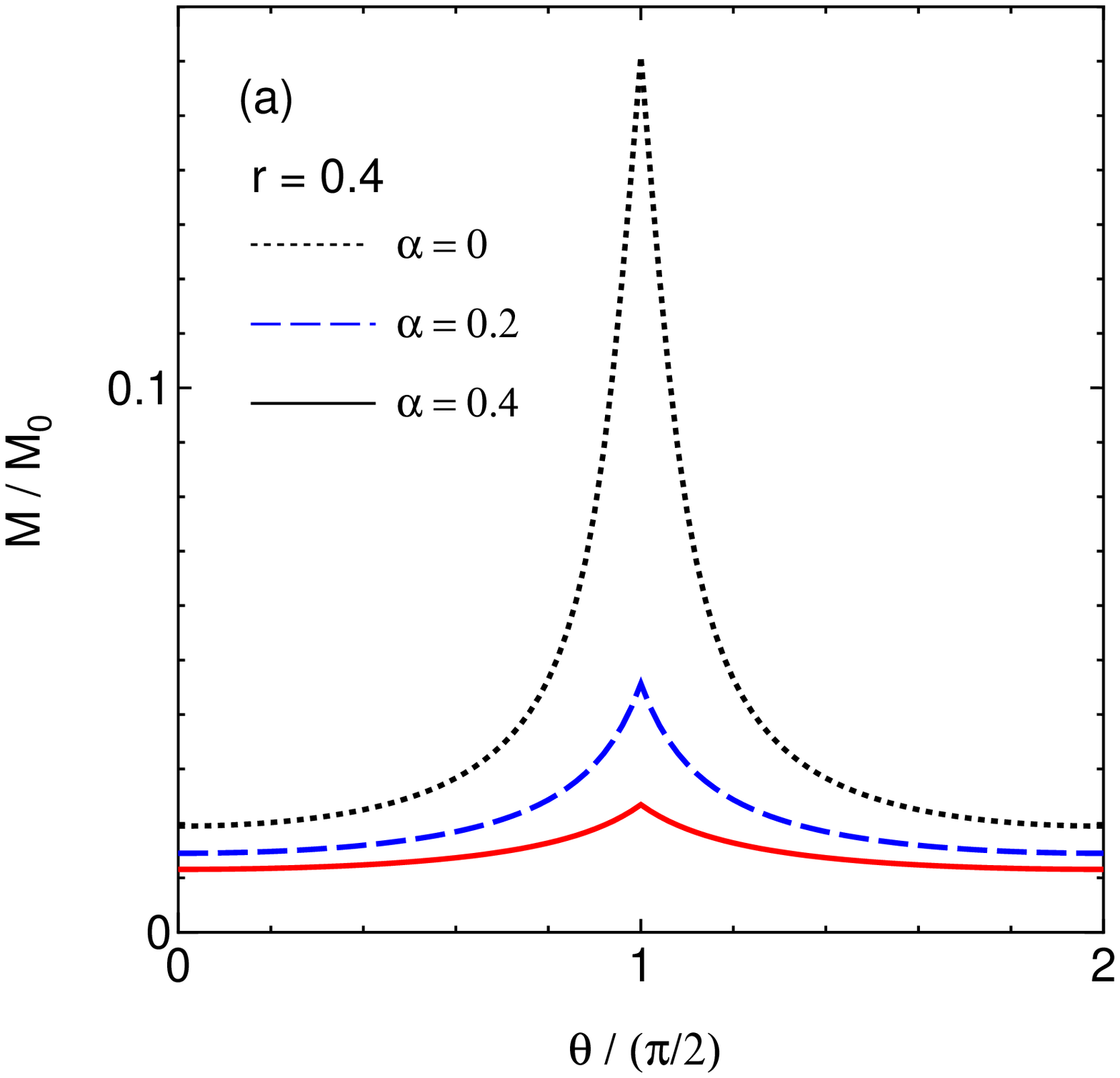}
\includegraphics[width=0.23\textwidth]{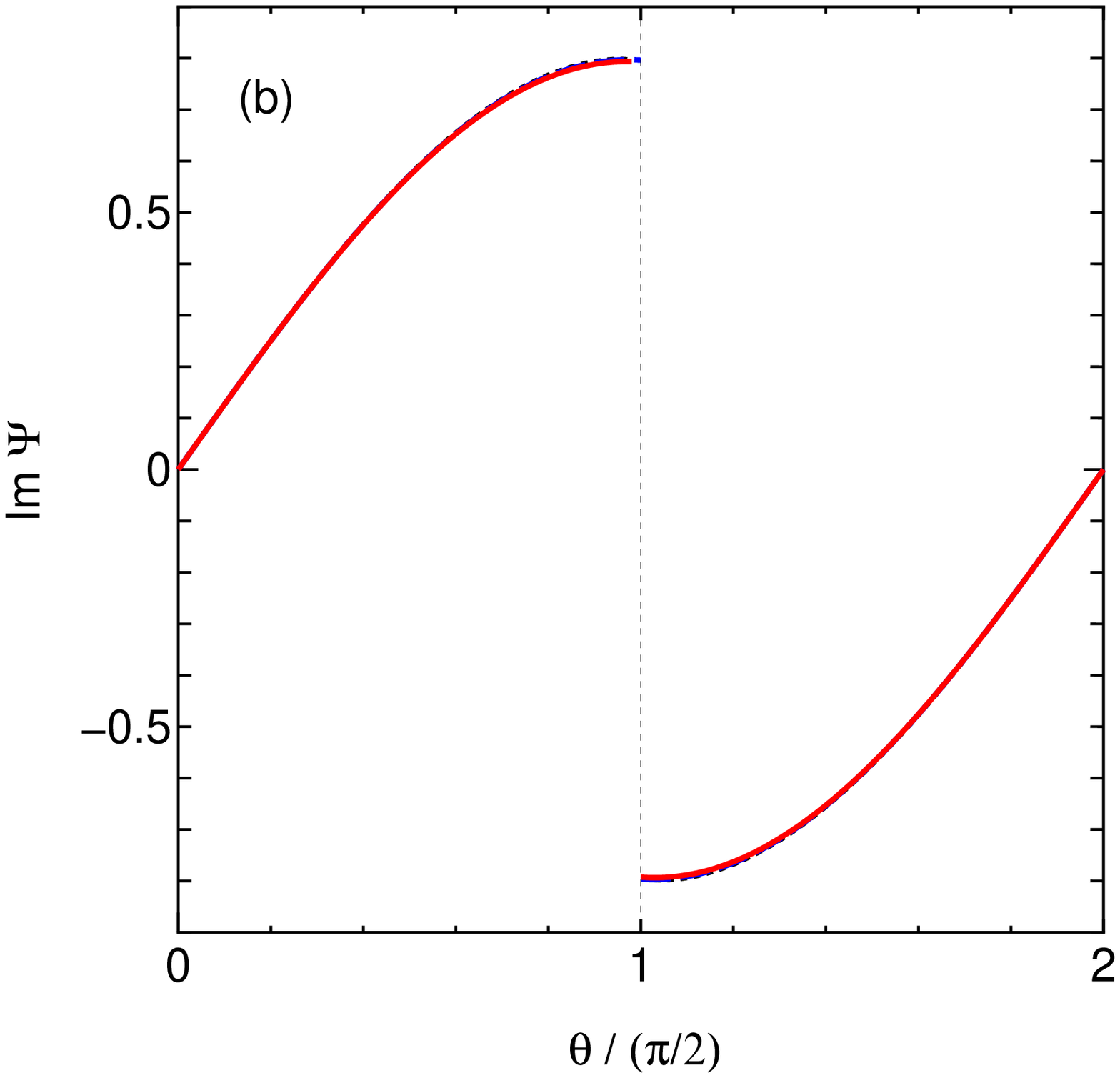}
\includegraphics[width=0.23\textwidth]{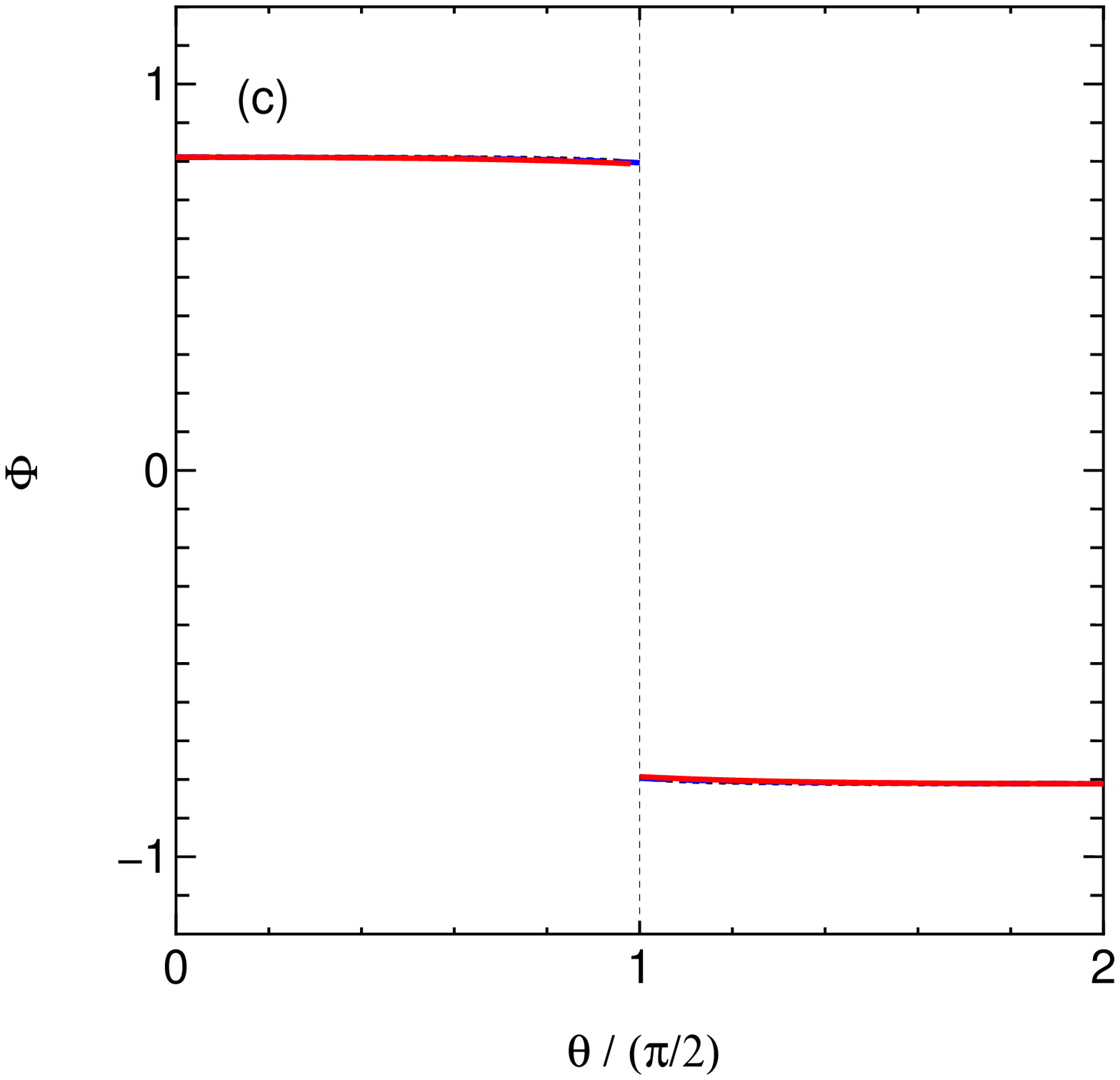}
\end{center}
\caption{$\theta$-dependence of order parameters
(a) $M$, (b) $\mathrm{Im}[\Psi]$ and (c) $\Phi$ at $T=2.5 m_\pi$. Here
$M$ is normalized by the value $M_0$ at $T=\mu=0$.
The dotted, dashed and solid lines represent the results of
$\alpha=0$, $0.2$ and $0.4$, respectively.
}
\label{Fig-OP-pio1}
\end{figure}

Figures \ref{Fig-OP-pio1}(a)-(c) represent $\theta$ dependence of $M$,
$\mathrm{Im}[\Psi]$ and $\Phi$, respectively, for three cases of
$\alpha=0$, $0.2$ and $0.4$.
Here we consider a high-$T$ case of $T=2.5 m_\pi$.
$M$ is an order parameter of chiral symmetry,
while $\mathrm{Im}[\Psi]$ and $\Phi$ are order parameters of ${\cal
C}{\mathbb Z}_{2}$ symmetry, respectively.
Equations \eqref{X-theta}-\eqref{Im-Phi-Z2} show that
$\Phi$ is antisymmetric with respect to the $\theta=\pi/2$ axis, whereas
$M$ is symmetric with respect to the axis.
In addition, $\mathrm{Im}[\Psi]$ is antisymmetric with respect to the
$\theta=\pi/2$ axis and zero at $\theta=0$ and $\pi$.
As shown in panel (b), $\mathrm{Im}[\Psi]$ has a gap at $\theta=\pi/2$,
indicating that ${\cal C}{\mathbb Z}_{2}$ symmetry is spontaneously
broken there for the high-$T$ case.
The zeroth-order discontinuity (gap) in $\mathrm{Im}[\Psi]$ means that a
first-order phase transition takes place at $\theta=\pi/2$.
This is the RW phase transition~\cite{Roberge:1986}.
At $\theta=\pi/2$, $\Phi$ also has a gap, since $\mathrm{Im}[\Psi]=\Phi$
at $\theta=\pi/2$.
This is a characteristic of two-color QCD.

The ${\cal C} {\mathbb Z}_2$-even quantity $M$ has a cusp at
$\theta=\pi/2$, when the ${\cal C} {\mathbb Z_2}$-odd quantities
$\mathrm{Im}[\Psi]$ and $\Phi$ have a gap there.
In general, the zeroth-order discontinuity (gap) in a ${\cal C} {\mathbb
Z_2}$-odd quantity is propagated to other ${\cal C} {\mathbb Z_2}$-odd
quantities as the zeroth-order discontinuity (gap) and to ${\cal C}
{\mathbb Z_2}$-even quantities as the first-order discontinuity
(cusp)~\cite{Barducci:1993od,Kashiwa:2009co}.
The order parameters $\mathrm{Im}[\Psi]$ and $\Phi$ are less sensitive
to $\alpha$ than $M$.

\subsection{Interplay between chiral and deconfinement transitions}
\label{Interplay}

\begin{figure}[htbp]
\begin{center}
\includegraphics[width=0.23\textwidth]{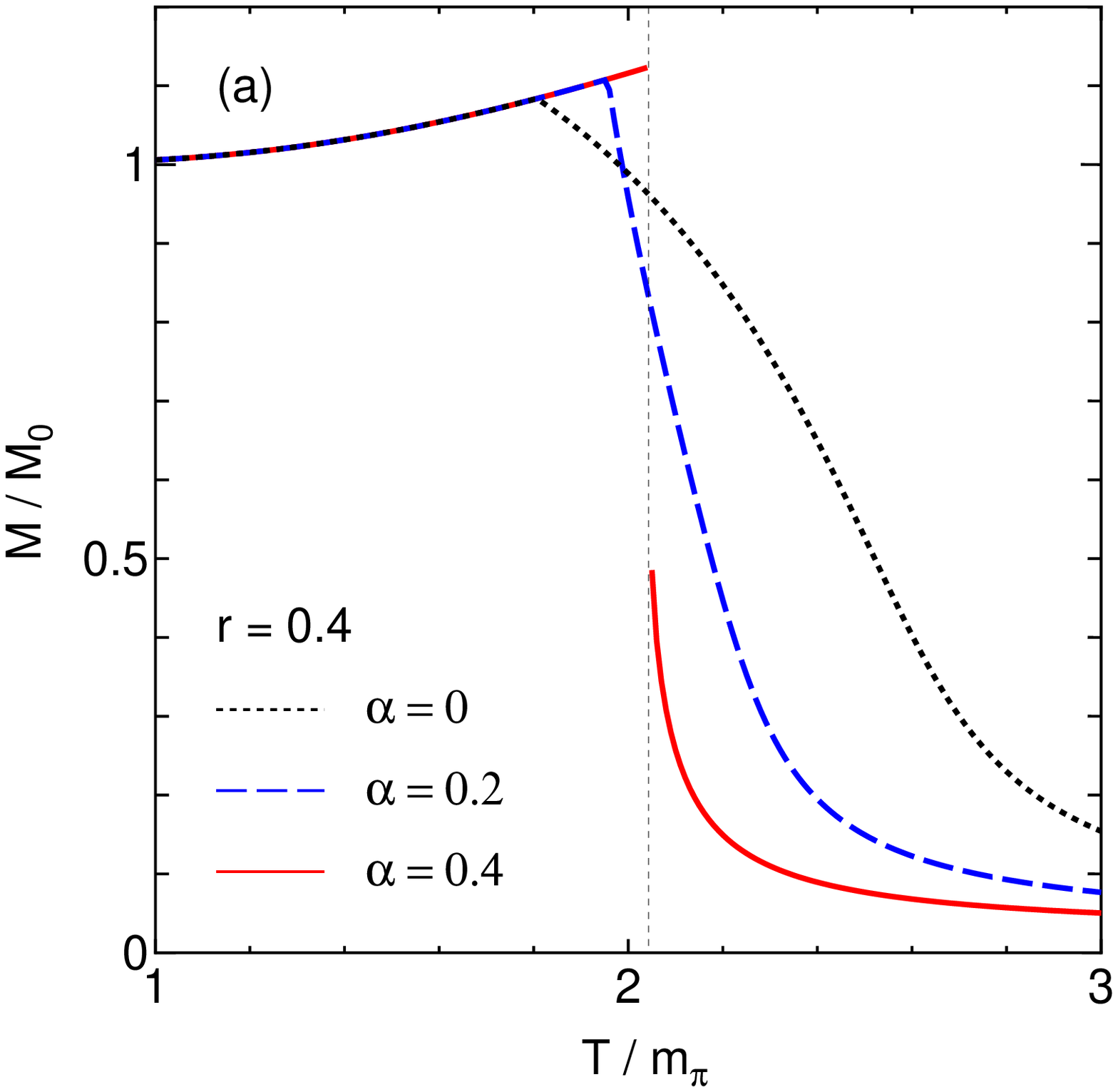}
\includegraphics[width=0.23\textwidth]{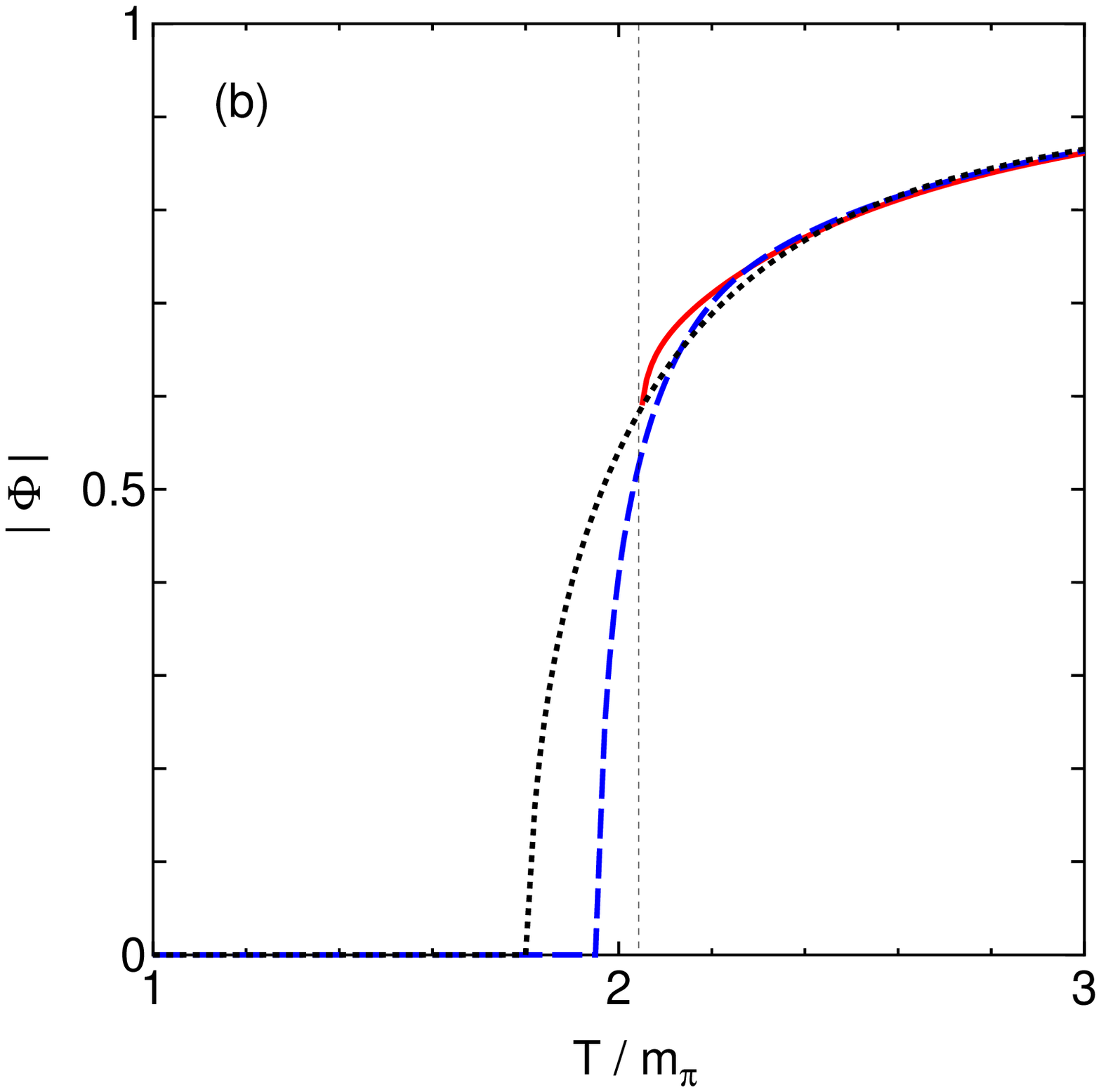}
\end{center}
\caption{$T$-dependence of (a) $M/M_0$ and (b) $|\Phi|$ at $\theta=\pi/2$.
The dotted, dashed and solid lines represent the results of
$\alpha=0$, $0.2$ and $0.4$, respectively.
}
\label{Fig-OP-pio2}
\end{figure}

Figure~\ref{Fig-OP-pio2} shows $T$-dependence of $M/M_0$ and
$|\Phi|$ at $\theta=\pi/2$ for three cases of $\alpha=0$, $0.2$ and $0.4$.
The ${\cal C}{\mathbb Z}_{2}$-symmetry breaking takes place at some
temperature $T^c_{\rm d}$, and the order is second-order for $\alpha=0$
and $0.2$, but becomes first-order for $\alpha=0.4$.
Here the vertical thin-dotted line denotes the critical temperature of
the first-order transition.
The chiral transition is crossover for $\alpha=0$ and 0.2, but it
becomes first-order for $\alpha=0.4$.
The entanglement thus intensifies both the chiral transition and the
${\cal C}{\mathbb Z}_{2}$-symmetry breaking.
For $\alpha=0$ and 0.2, $M$ has a cusp at $T=T^{c}_{\rm d}$ as a result
of the propagation of the first-order discontinuity in
$|\Phi|$~\cite{Kashiwa:2009co}.

\begin{figure}[htbp]
\begin{center}
\includegraphics[width=0.23\textwidth]{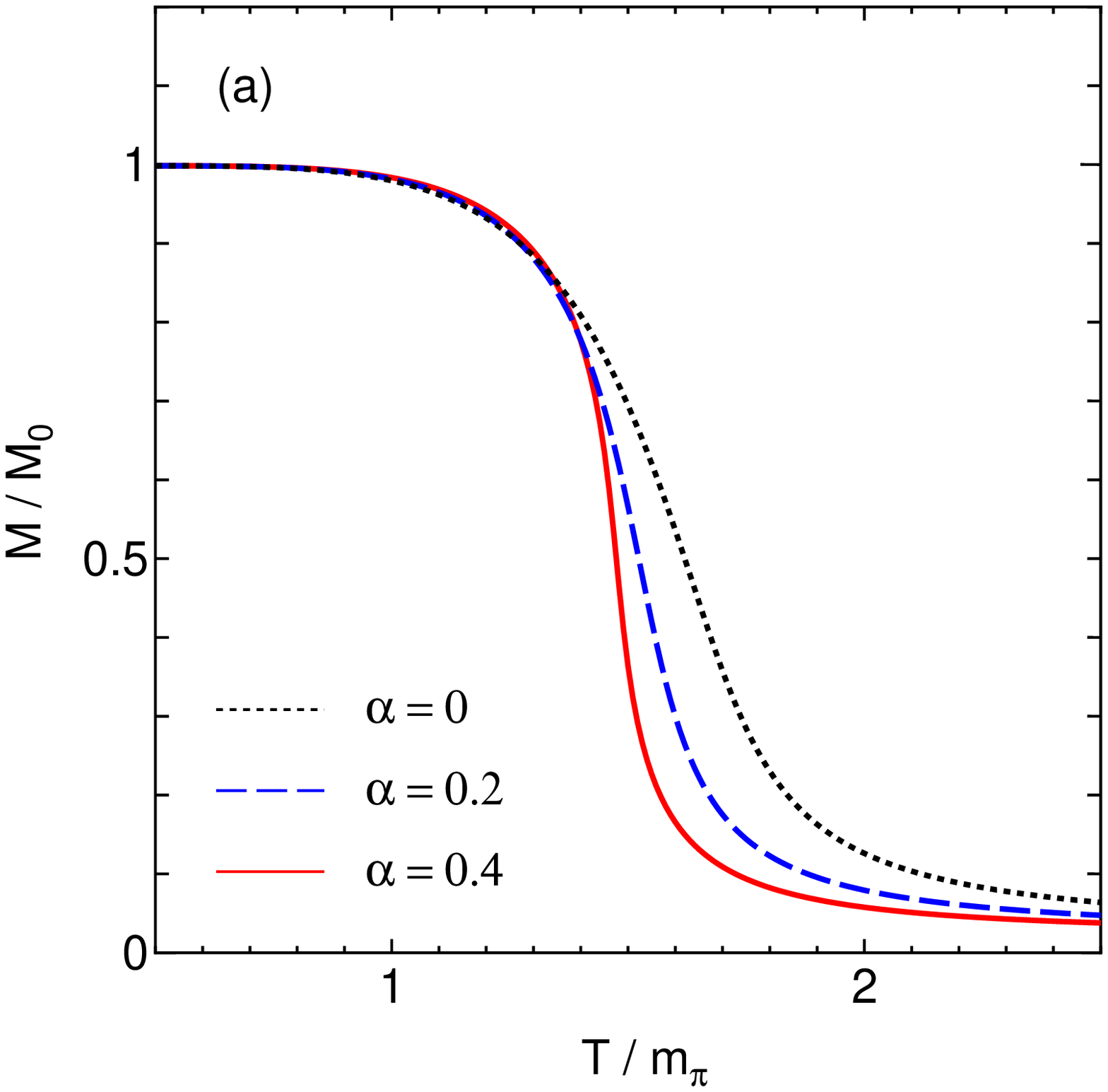}
\includegraphics[width=0.23\textwidth]{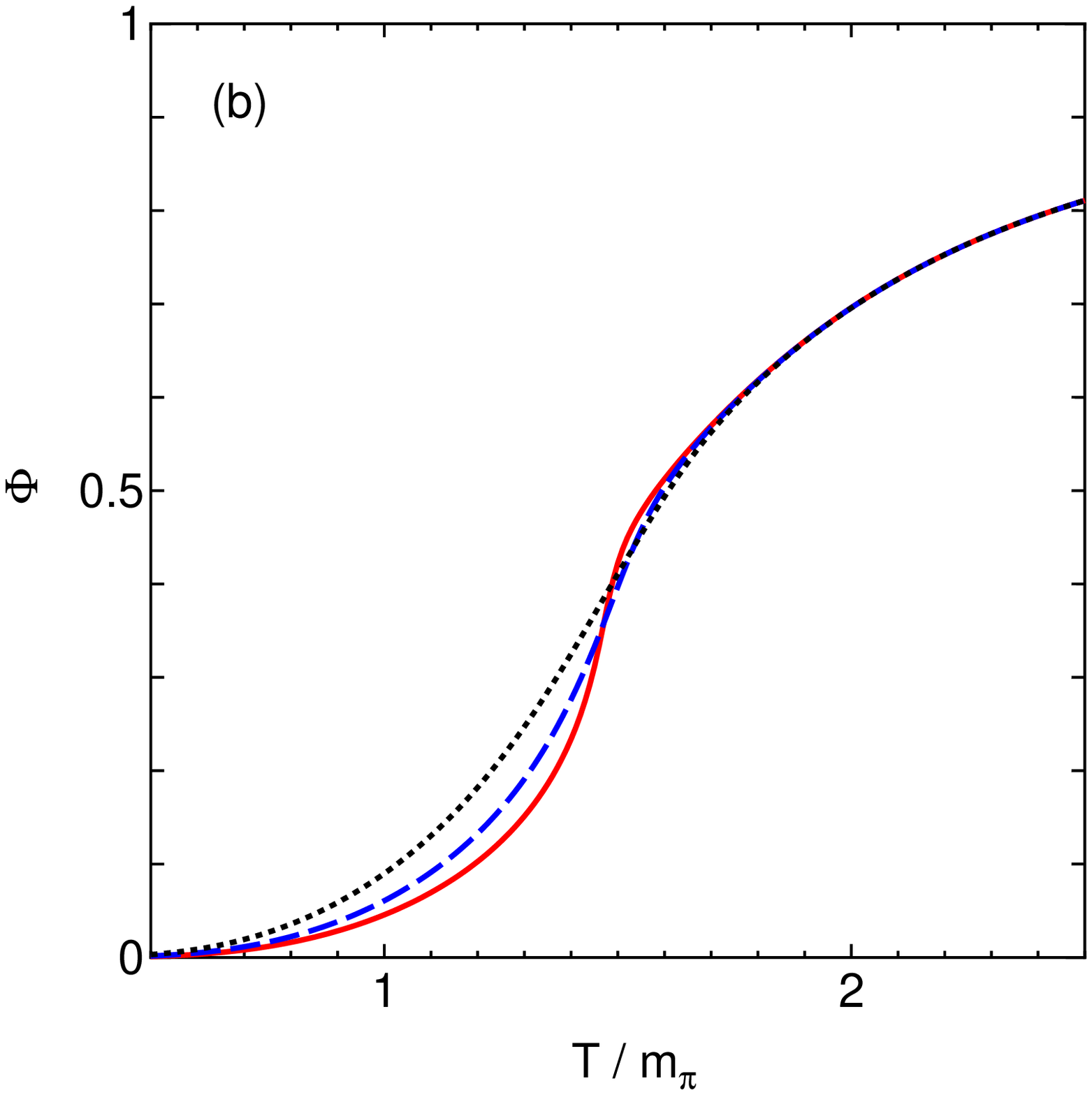}
\end{center}
\caption{$T$-dependence of (a) $M/M_0$ and (b) $\Phi$ at $\mu=0$.
See Fig.~\ref{Fig-OP-pio2} for the definition of lines.
}
\label{Fig-OP-mu0}
\end{figure}

Figure~\ref{Fig-OP-mu0} shows $T$-dependence of $M/M_0$ and $\Phi$ at
$\theta=0$ ($\mu=0$), where $\Phi$ is an order parameter of the
deconfinement transition.
Both the deconfinement and chiral transitions keep crossover for
$\alpha=0$, $0.2$ and $0.4$, although the transitions become stronger as
$\alpha$ increases.
The pseudo-critical temperature $T^{c}_{\rm d}$ of the deconfinement
transition is less sensitive to $\alpha$ than the pseudo-critical
temperature
$T^{c}_{\chi}$ of the chiral transition, where the pseudo-critical
temperatures are defined by peak positions of $dM/dT$ and $d\Phi/dT$,
respectively.

As shown in Fig \ref{Fig-OP-pio2}, the ${\cal C}{\mathbb Z}_{2}$
symmetry breaking takes place on a line of $\theta=\pi/2$ and $T \ge
T^c_{\rm d}$.
As discussed in Sec. \ref{Formalism}, the line is the RW phase
transition line and the endpoint is nothing but the endpoint of the RW
phase transition line; therefore, $T^c_{\rm d}=T^c_{\rm RW}$.
This endpoint is called the RW endpoint.
As shown later in Fig. \ref{PD-I}, the spontaneous breaking of ${\cal
C}{\mathbb Z}_{2}$ symmetry at the RW endpoint is continuously connected
to the deconfinement transition at $0 \le \theta <\pi/2$.
Thus the crossover deconfinement transition at $\theta=0$ is a remnant
of the ${\cal C}{\mathbb Z}_{2}$ symmetry breaking at the RW endpoint.
Hence, for simplicity, we regard the ${\cal C}{\mathbb Z}_{2}$ symmetry
breaking at the RW endpoint as a part of the deconfinement transition at
$0 \le \theta < \pi/2$.

Dynamics at $\theta=\pi/2$ is more complicated than that at $\mu=0$,
since the order parameters have discontinuities of either the zeroth or
the first order at $\theta=\pi/2$.
The order parameters are considered to be more sensitive to $\alpha$
near the discontinuities.
This is really true as seen by comparing Fig.~\ref{Fig-OP-pio2} with
Fig.~\ref{Fig-OP-mu0}.
Through the discontinuities, we can then investigate clearly how strong
the entanglement between the chiral and deconfinement transitions is.
The symmetry breakings at $\theta=\pi/2$ thus give deeper understanding.

\begin{figure}[htbp]
\begin{center}
\includegraphics[width=0.23\textwidth]{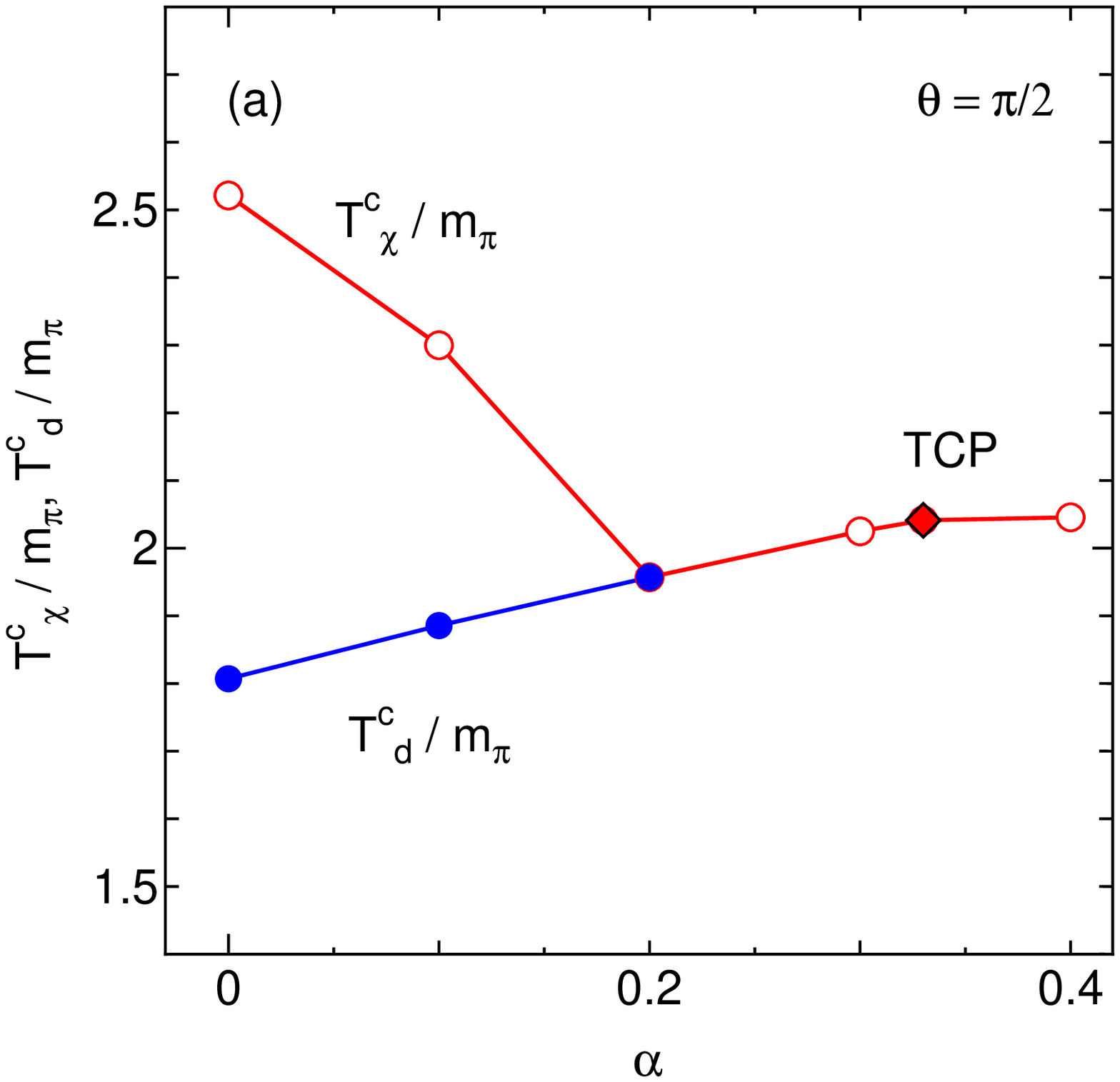}
\includegraphics[width=0.23\textwidth]{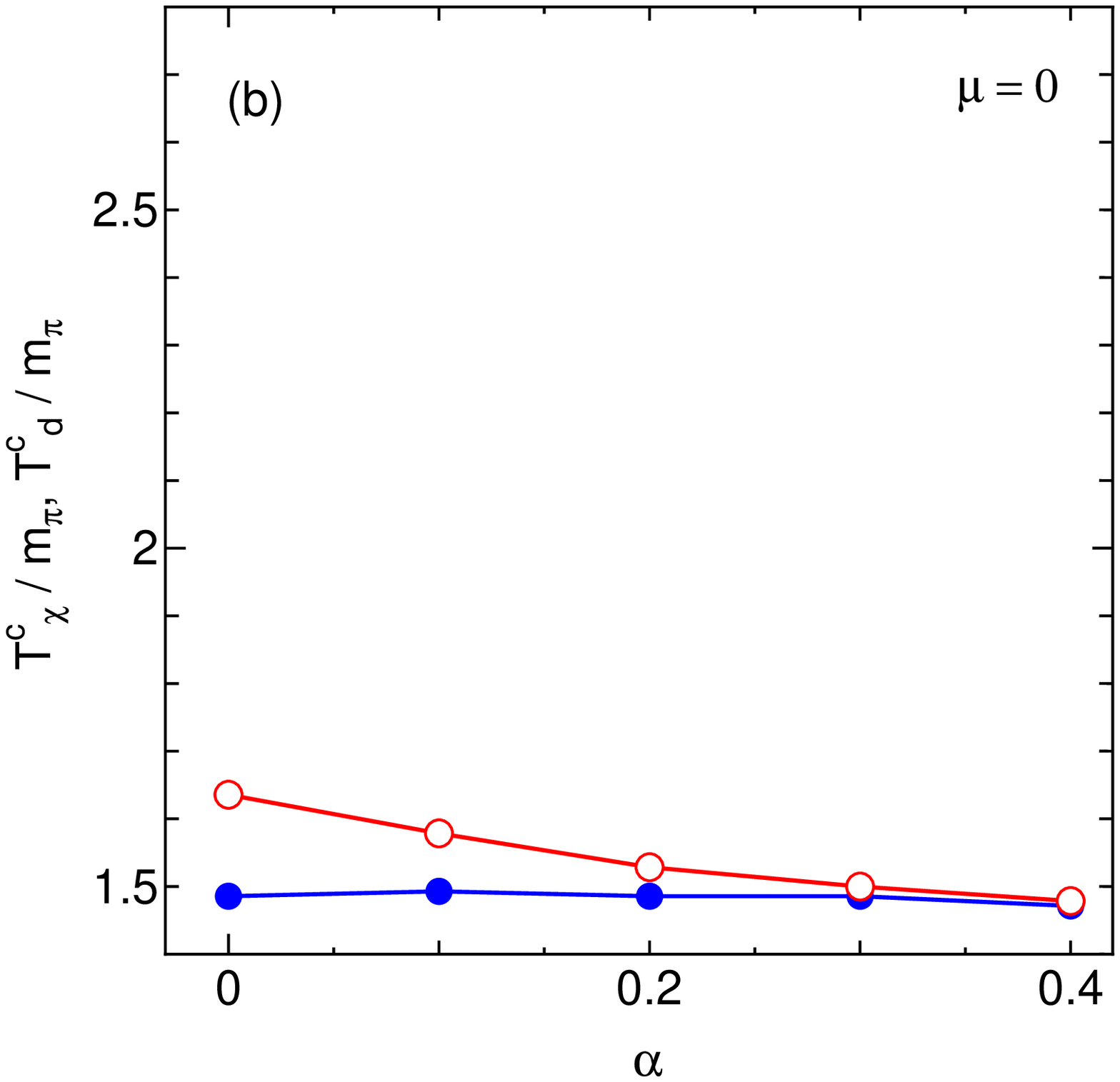}
\end{center}
\caption{
$\alpha$-dependence of (pseudo)critical temperatures
$T^{c}_{\rm d}$ and $T^{c}_{\chi}$ at (a) $\theta=\pi/2$ and (b) $\mu=0$
for the case of $G_\mathrm{v}=0.4 G$.
The close diamond symbol with ``TCP/CEP" stands for the point that is a
TCP for the deconfinment transition and a CEP for the chiral transition.
The numerical values have ambiguities of about 1 MeV and the solid lines
are drawn by connecting two neighborhood points.
}
\label{Fig-PCT}
\end{figure}

Figure~\ref{Fig-PCT} shows $\alpha$-dependence of two (pseudo)critical
temperatures $T^{c}_{\rm d}$ and $T^{c}_{\chi}$ at $\theta=\pi/2$ and $0$.
The two (pseudo)critical temperatures approach each other as $\alpha$
increases, and finally agree with each other at $\alpha \ga 0.2$ for
$\theta=\pi/2$ and at $\alpha \ga 0.4$ for $\theta=0$.
As for $\alpha \lessa 0.2$, the speed of the approach is much faster for
$\theta=\pi/2$ than for $\theta=0$.
The entanglement thus makes it stronger the correlation between the
chiral and deconfinement transitions.
Therefore, the difference $|T^{c}_{\chi}-T^{c}_{\rm d}|$ is a good
quantity to determine the value of $\alpha$.
For $\theta=\pi/2$, the deconfinement transition is second-order at
$\alpha < \alpha_c \approx 0.33$, but first-order at $\alpha >
\alpha_c$.
The chiral transition is, meanwhile, crossover at $\alpha < \alpha_c$,
although it is first-order at $\alpha > \alpha_c$.
Hence, the point at $\alpha = \alpha_c$ is a tri-critical point (TCP)
for the deconfinement transition and a critical endpoint for the chiral
transition.

\begin{figure}[htbp]
\begin{center}
\includegraphics[width=0.23\textwidth]{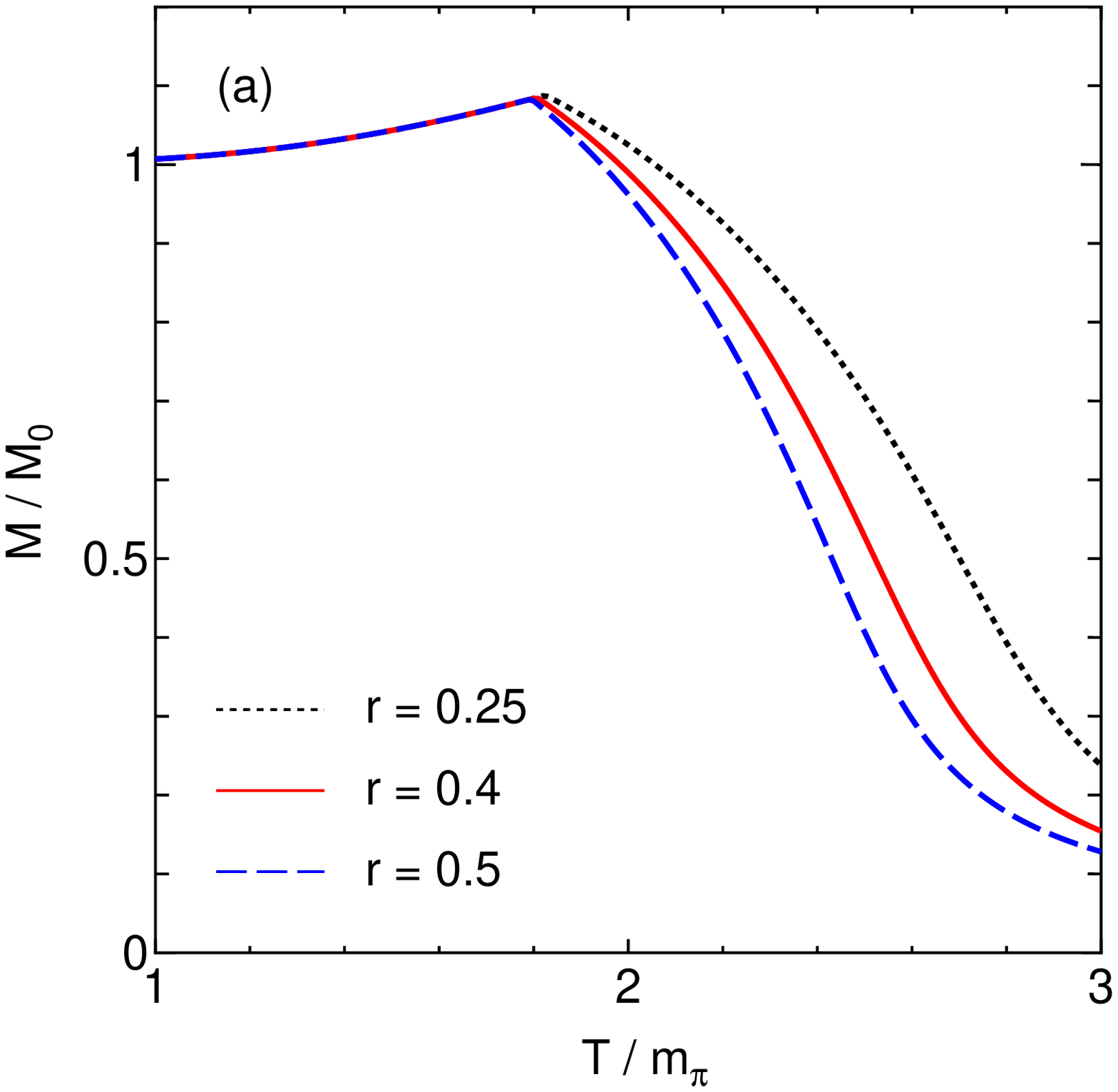}
\includegraphics[width=0.23\textwidth]{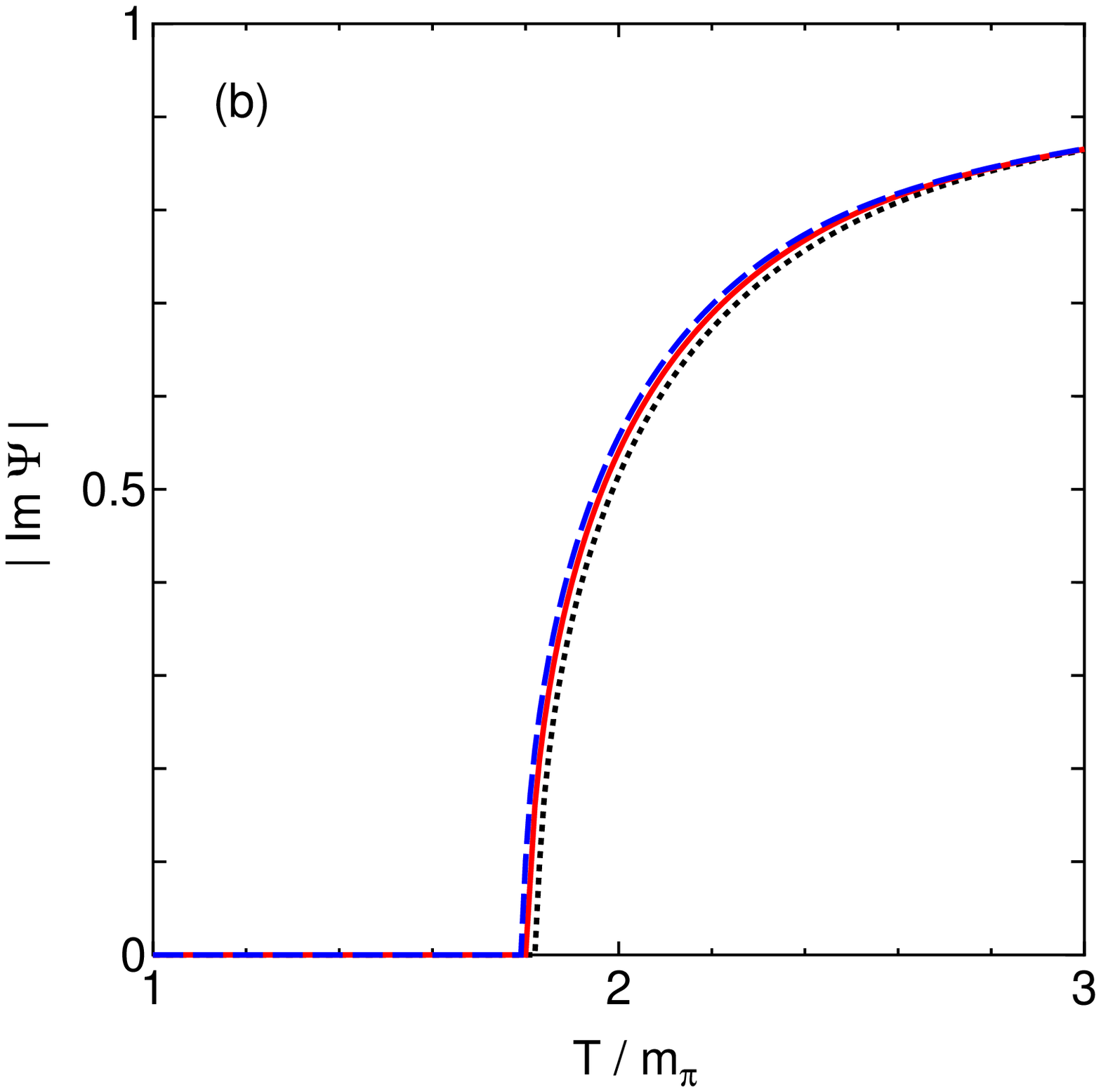}
\end{center}
\caption{
$T$-dependence of (a) $M/M_0$ and (b) $|\mathrm{Im}~\Psi|$ at
$\theta=\pi/2$.
The dotted, dashed and solid lines denote the results for $r=0.25$,
$0.4$ and $0.5$, respectively. Here $\alpha$ is set to $0.4$.
}
\label{Fig:v-dep}
\end{figure}

Figure~\ref{Fig:v-dep} shows $r$-dependence of $M/M_0$ and
$|\mathrm{Im}~\Psi|$ at $\theta=\pi/2$.
The effect of $r$ is similar to that of $\alpha$, but the former effect
is smaller than the latter one, when $r$ is varied within a realistic
range from 0.25 to 0.5.

\subsection{Phase diagram at imaginary and real chemical potentials}
\label{Phase diagram}

\begin{figure}[htbp]
\begin{center}
\includegraphics[width=0.23\textwidth]{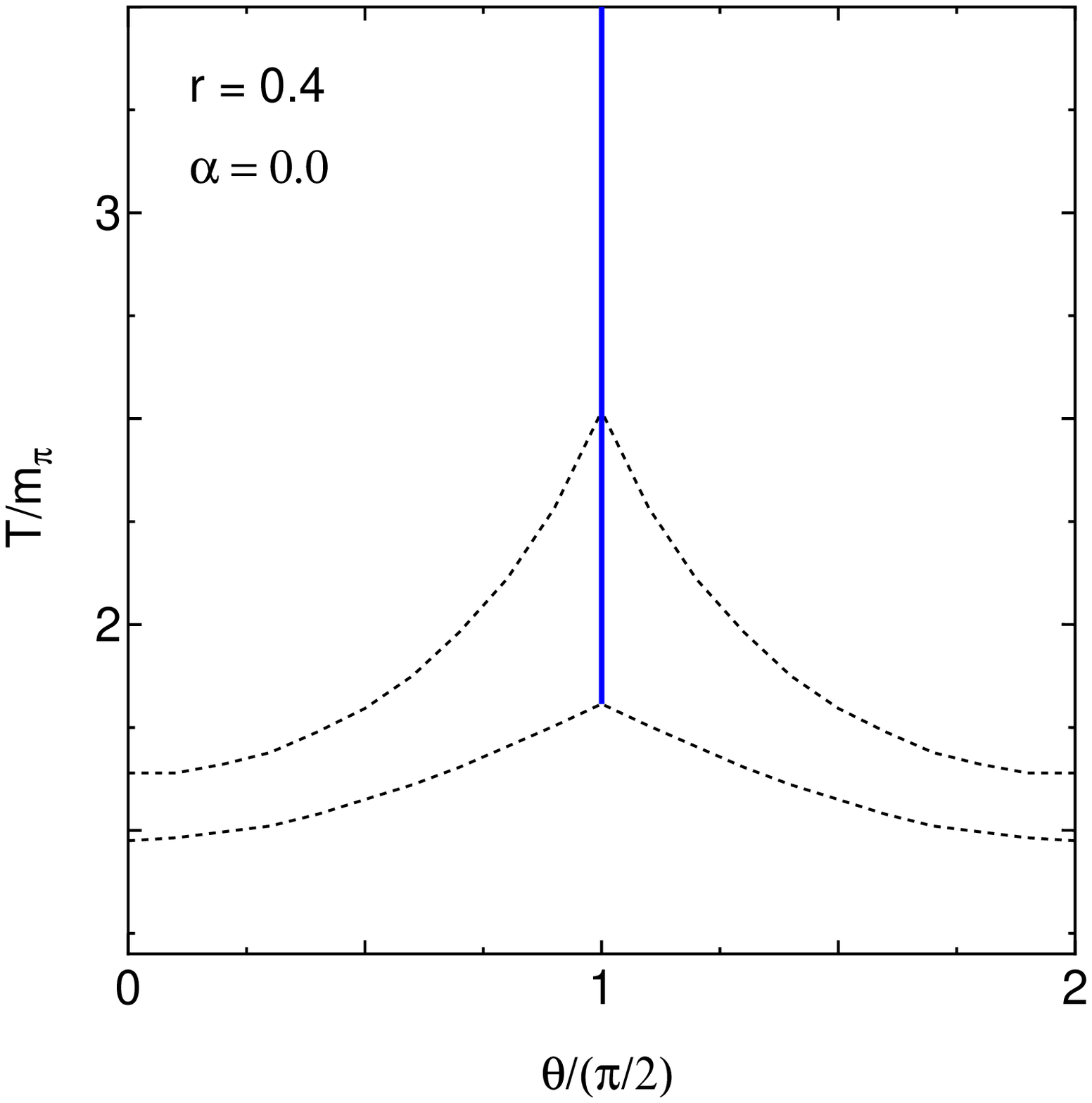}
\includegraphics[width=0.23\textwidth]{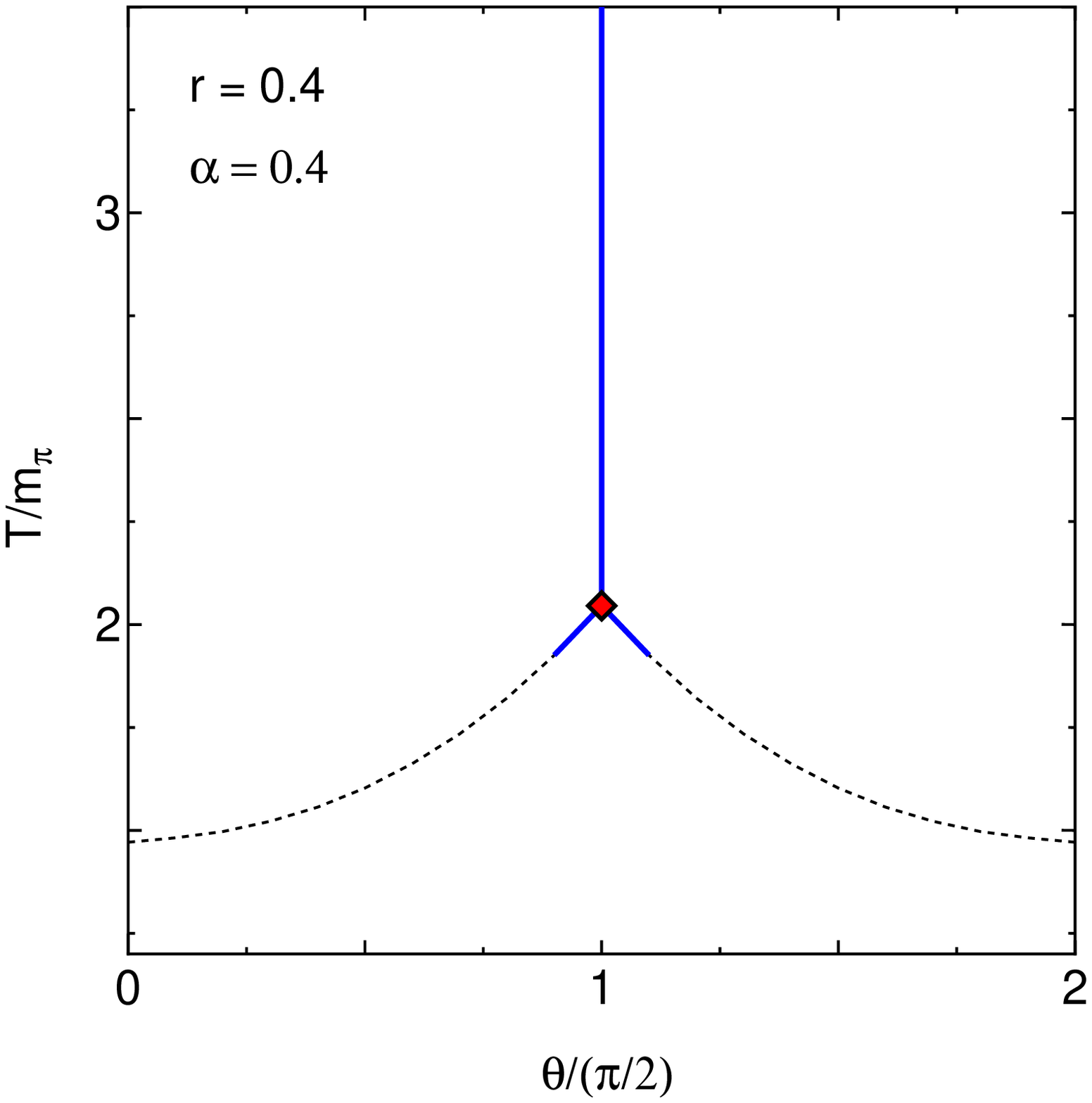}
\end{center}
\caption{
The phase diagram in the $\theta$-$T$ plane for $\alpha=0$ (left panel)
and $\alpha=0.4$ (right panel).
Here the case of $r=0.4$ is taken.
The dotted and solid lines stand for crossover and first-order
transitions, respectively.
In the left panel, the upper and lower dotted lines mean the chiral and
deconfinement crossover lines, respectively.
In the right panel, the chiral and deconfinement crossover lines almost
coincide with each other, and the diamond symbol denotes the triple-point.
}
\label{PD-I}
\end{figure}

First we consider the phase diagram at imaginary $\mu$ for two cases of
$\alpha=0$ and 0.4. Here we take $r=0.4$.
Figure \ref{PD-I} shows the phase diagram in the $\theta$-$T$ plane.
In the left panel for the case of $\alpha=0$, the upper and lower dotted
lines denote chiral and deconfinement crossover transitions,
respectively. The two transitions are thus separated from each other,
when the correlation between the two transitions is weak.
In this situation, the ${\cal C}{\mathbb Z}_2$ symmetry breaking at the
RW endpoint $(\theta,T)=(\pi/2,T^c_{\rm RW})$ is second-order.
In the right panel for the strong correlation case of $\alpha=0.4$, the
two crossover transitions (dotted lines) agree with each other and the
${\cal C}{\mathbb Z}_2$ symmetry breaking at the RW endpoint becomes
first-order.
In other words, the RW endpoint becomes a triple-point where three
first-order transition lines meet.
Thus, the RW endpoint becomes a triple-point, when the correlation
between the chiral and deconfinement transition are strong.

The Polyakov-loop effective potential used in this study yields the
second-order
${\mathbb Z}_2$ symmetry breaking in the heavy quark limit.
Since the entanglement parameter $\alpha$ makes the symmetry breaking
stronger, the first-order ${\cal C}{\mathbb Z}_2$ symmetry breaking in
the physical quark mass does not come from the ${\cal C}{\mathbb Z}_2$
symmetry breaking only.
The chiral transition becomes stronger, if the coupling constant
$G$ is weakened at finite $T$; see for example Ref.~\cite{Kashiwa:2007}.
In the present model, the coupling constant is weakened through the
entanglement vertex, so that the chiral transition becomes first-order.
This behavior is seen in the left panel of Fig.~\ref{Fig-OP-pio2}.
The first-order ${\cal C}{\mathbb Z}_2$ symmetry breaking is therefore
induced by the first-order chiral transition.

\begin{figure}[htbp]
\begin{center}
\includegraphics[width=0.23\textwidth]{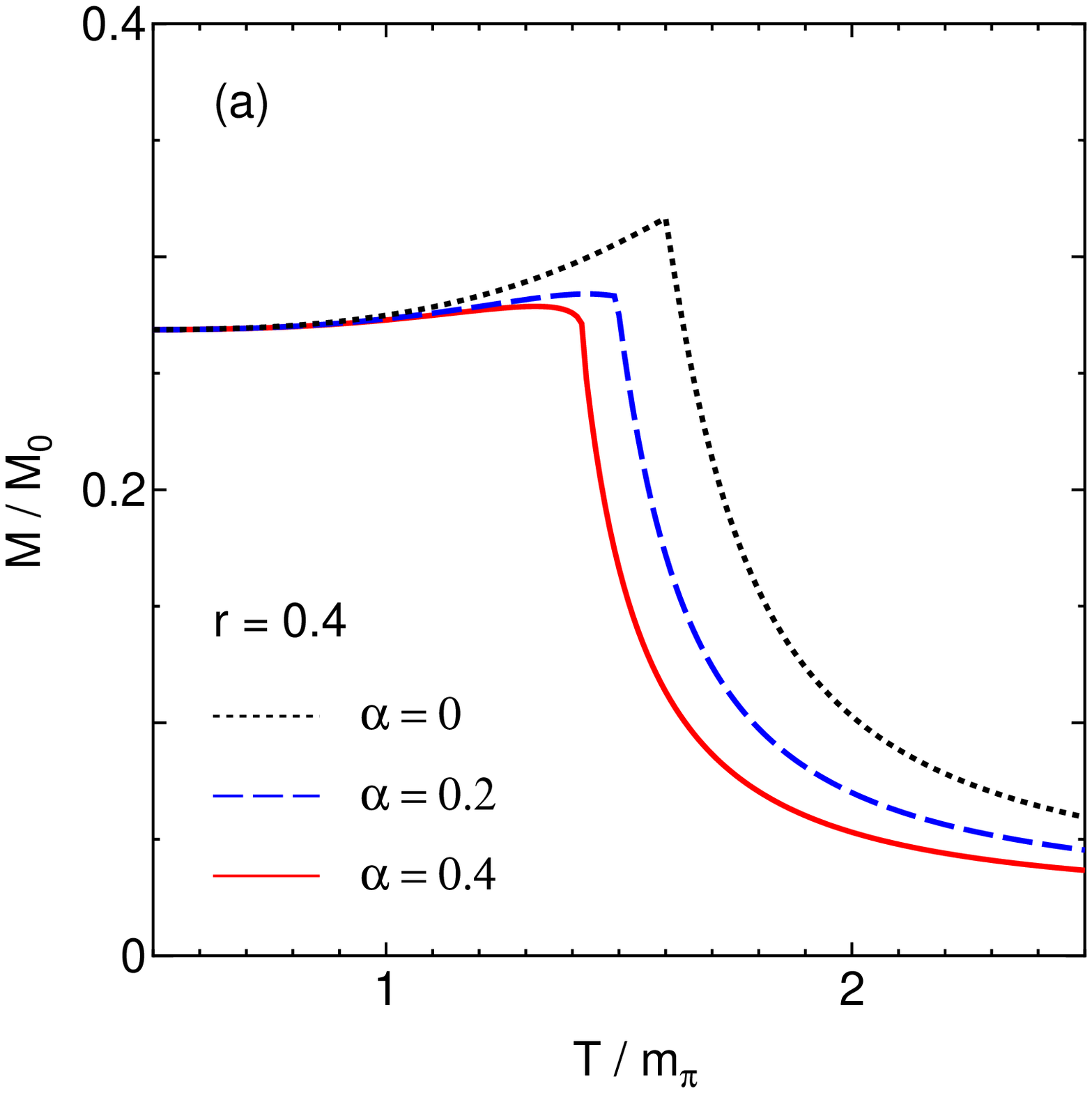}
\includegraphics[width=0.23\textwidth]{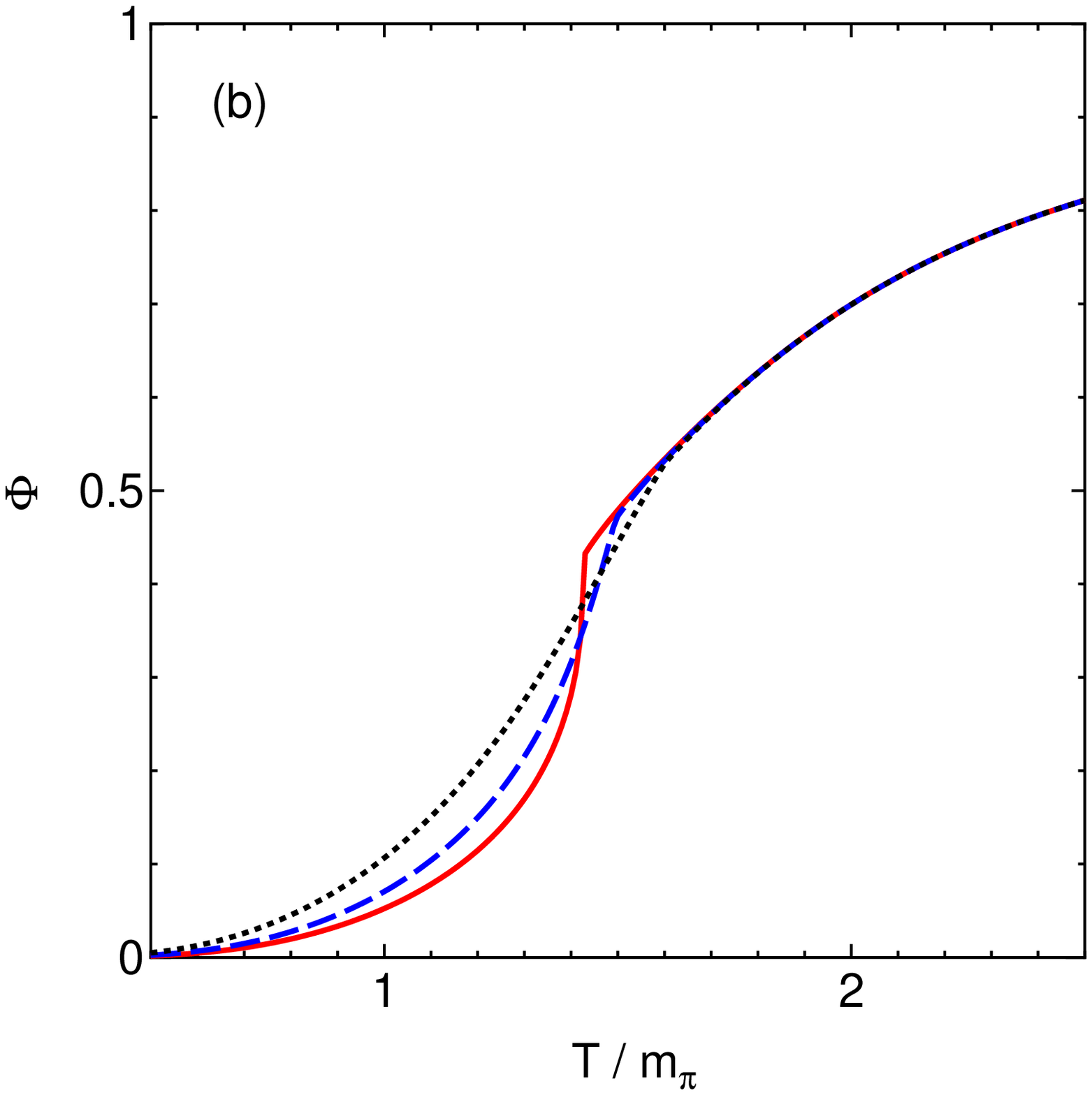}
\includegraphics[width=0.23\textwidth]{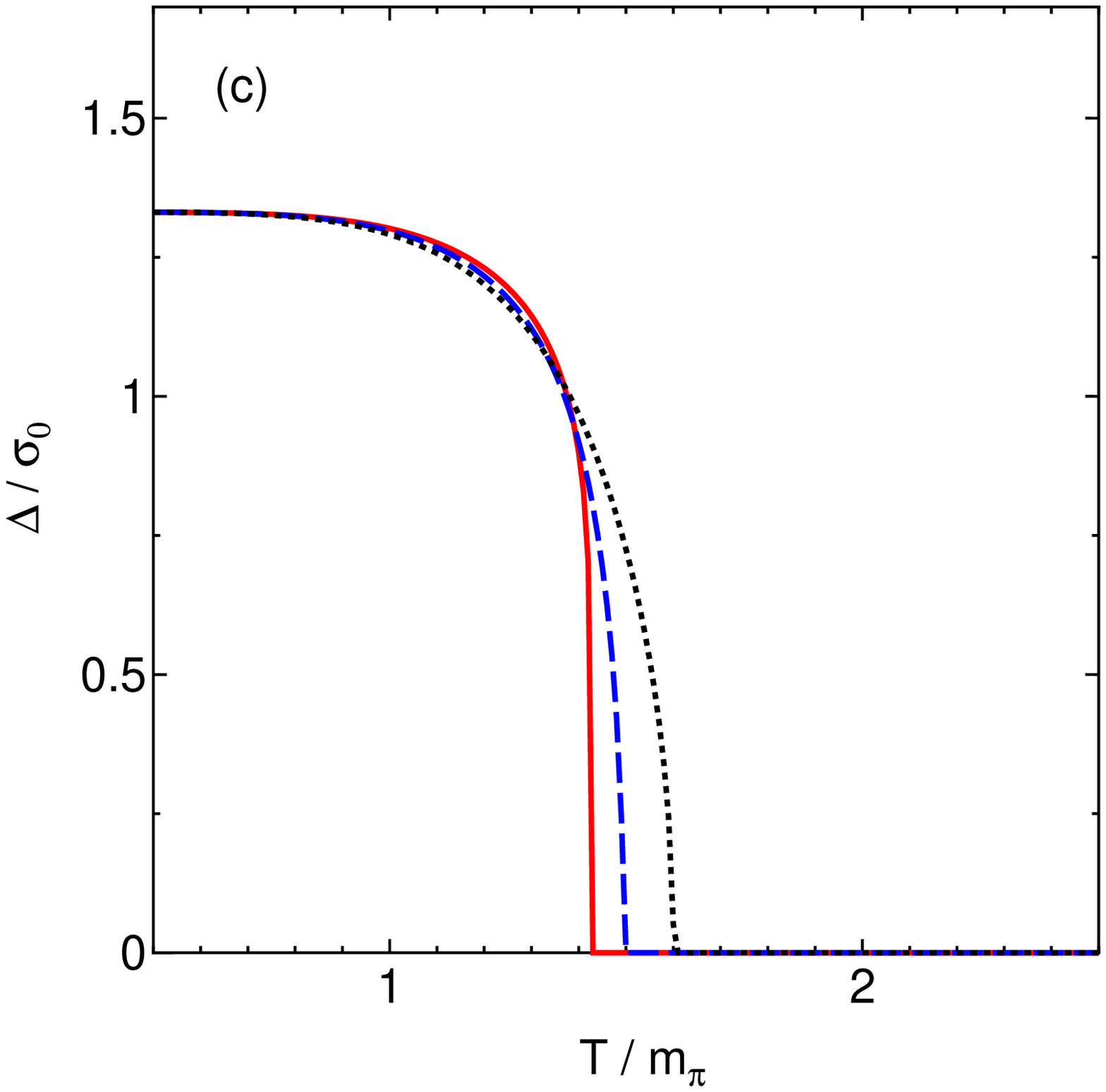}
\includegraphics[width=0.23\textwidth]{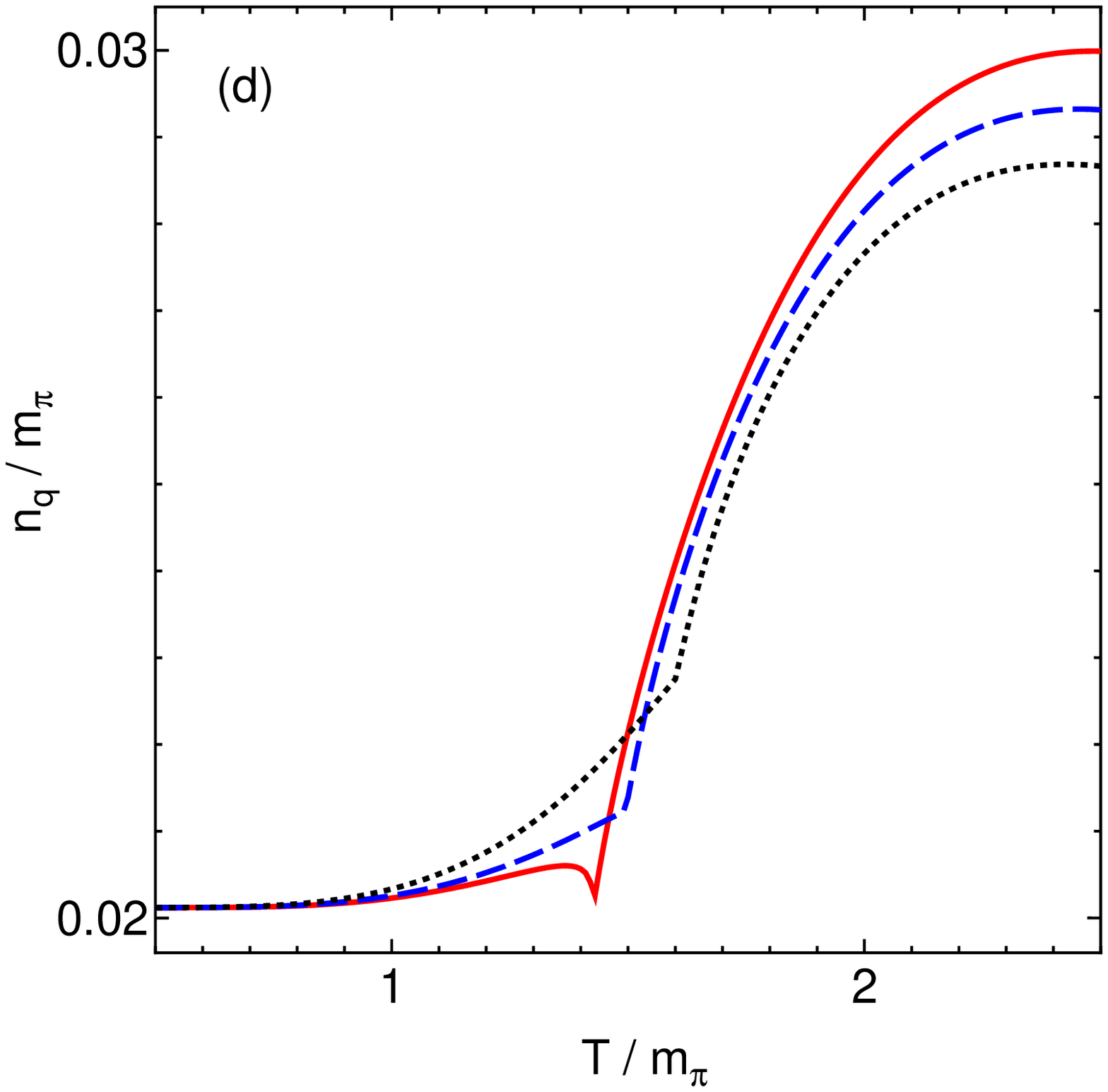}
\end{center}
\caption{$T$-dependence of (a) $M/M_0$,
(b) $\Phi$, (c) $\Delta/\sigma_0$ and (d) $n_q$ for the case of
$r=0.4$ and $\mu=m_\pi$.
The dotted and solid lines represent the results of $\alpha = 0$ and
$0.4$, respectively.
}
\label{T-muR}
\end{figure}

Next we investigate an influence of $\alpha$ on the phase diagram at
real $\mu$. Figure \ref{T-muR} shows $T$-dependence of $M/M_0$,
$\Phi$, $\Delta / \sigma_0$ and $n_q$ for the case of $r=0.4$ and
$\mu=m_{\pi}$, where $\sigma_0$ is the absolute value of $\sigma$ at
$T=\mu=0$.
As shown in panel (c), $\Delta / \sigma_0$ vanishes around
$T=1.5m_{\pi}$, indicating that the superfluid/normal transition occurs
there.
The phase boundary is rather sensitive to $\alpha$.

\begin{figure}[htbp]
\begin{center}
\includegraphics[width=0.23\textwidth]{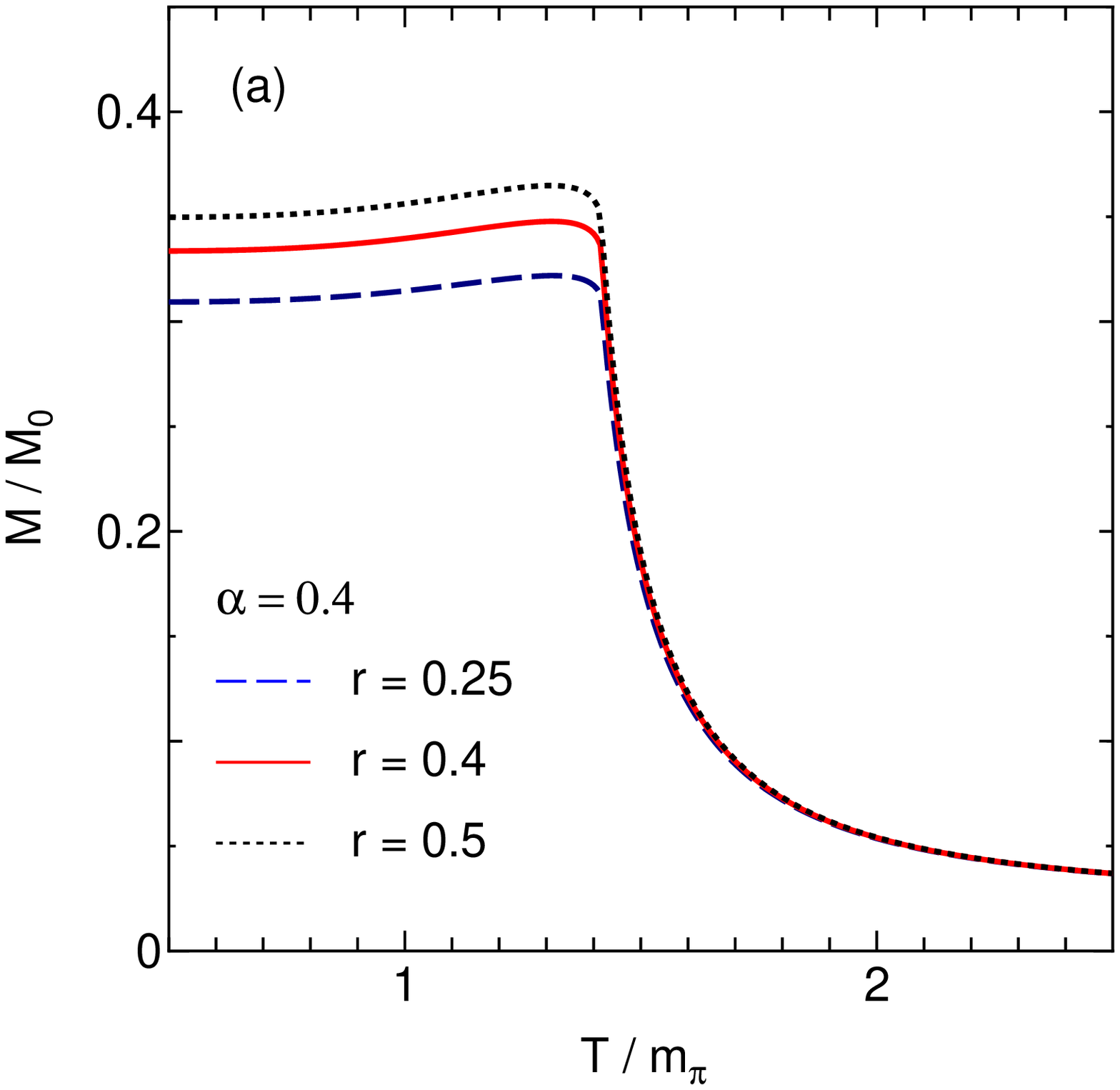}
\includegraphics[width=0.23\textwidth]{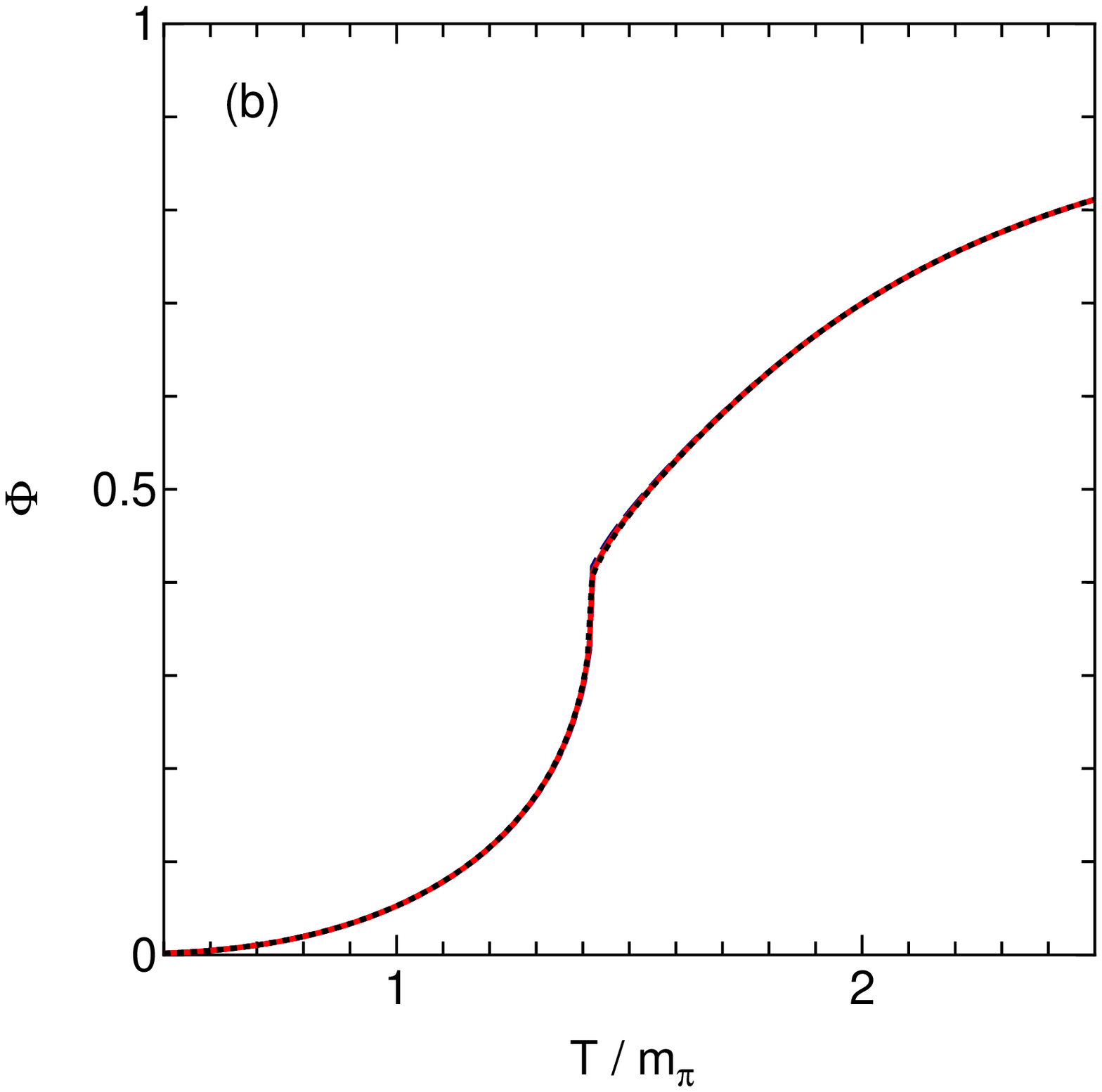}
\includegraphics[width=0.23\textwidth]{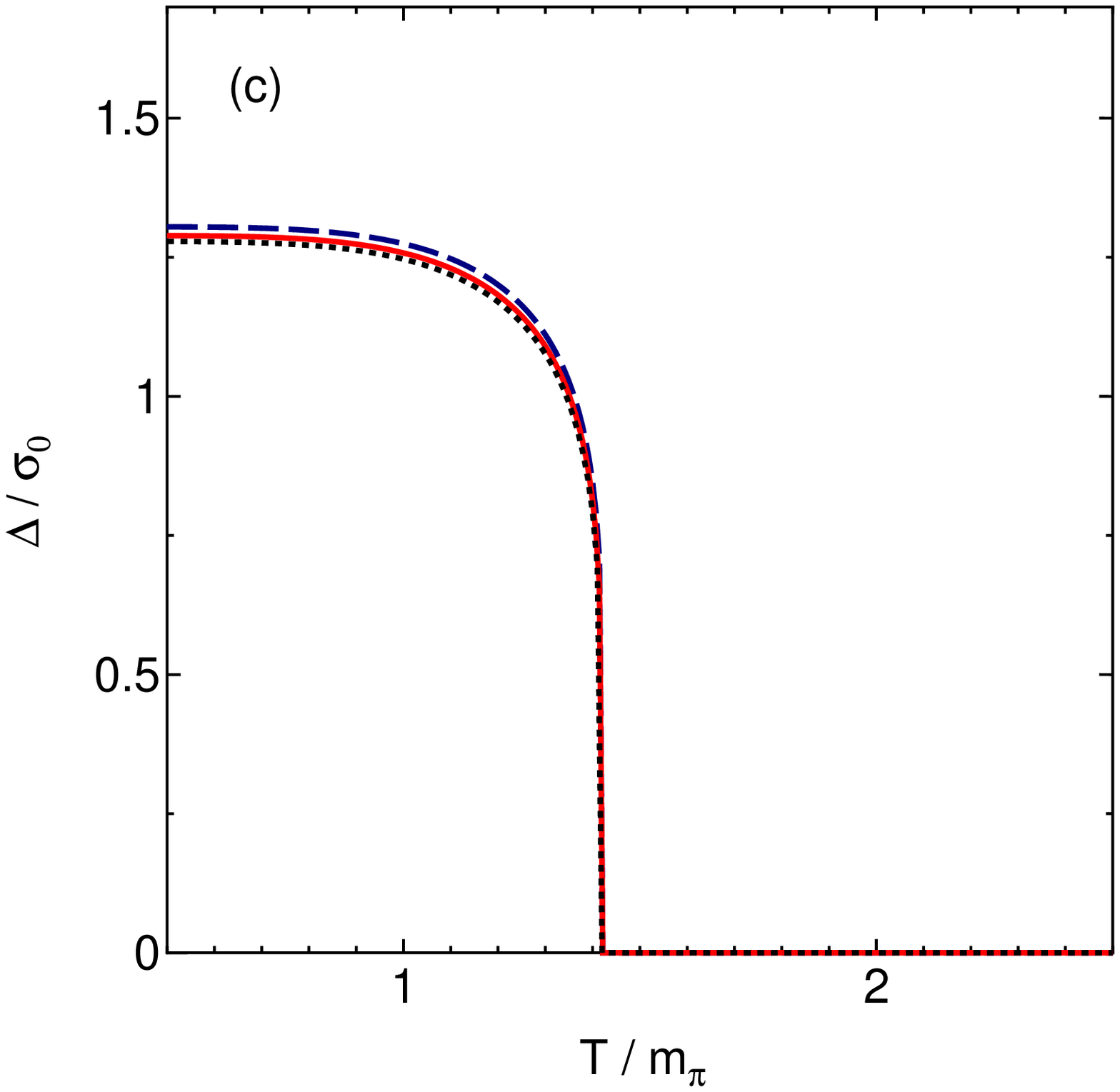}
\includegraphics[width=0.23\textwidth]{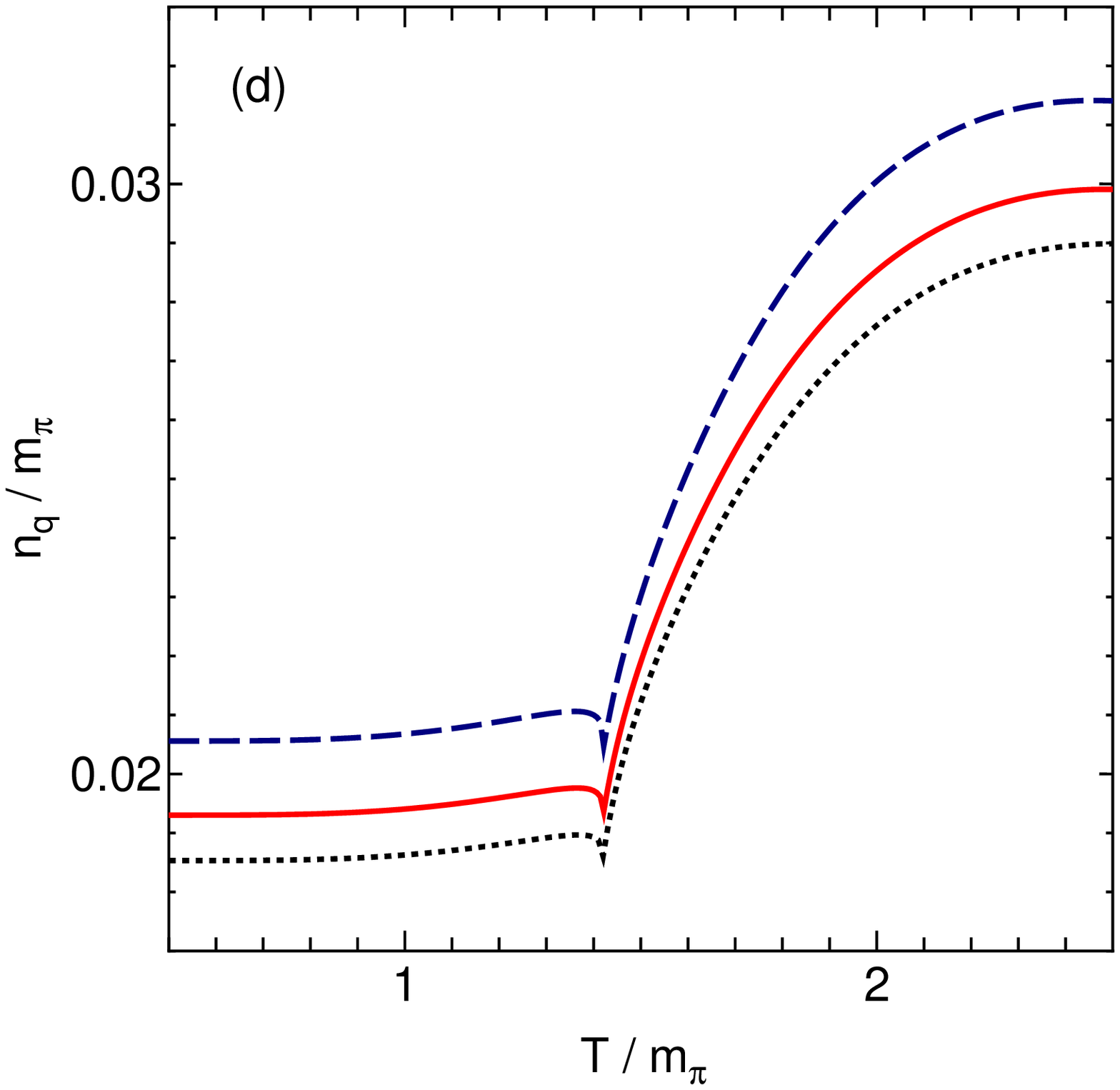}
\end{center}
\caption{$T$-dependence of (a) $M/M_0$,
(b) $\Phi$, (c) $\Delta/\sigma_0$ and (d) $n_q$ for the case of
$\alpha=0.4$ and $\mu=m_\pi$.
The dotted, solid and dashed lines represent the results of $r = 0.25$,
$0.4$ and $0.5$, respectively.
}
\label{T-muR-v}
\end{figure}

In this study, we take $r \equiv G_{\rm v}/G =0.4$ for all the calculations.
Although this value is obtained from the $N_c=3$ case by comparing the
PNJL results with the corresponding LQCD data, it is not easy to
determine the value definitely.
We then check $r$-dependence of order parameters, as shown in
Fig.~\ref{T-muR-v} where $r = 0.25$ and $0.5$ are taken as lower and
upper limits of a reliable range of $r$.
Effects of $r$ on $M$, $\Phi$ and $\Delta$ are rather small, although
$r$ gives an appreciable effect on $n_q$.

\begin{figure}[htbp]
\begin{center}
\includegraphics[width=0.23\textwidth]{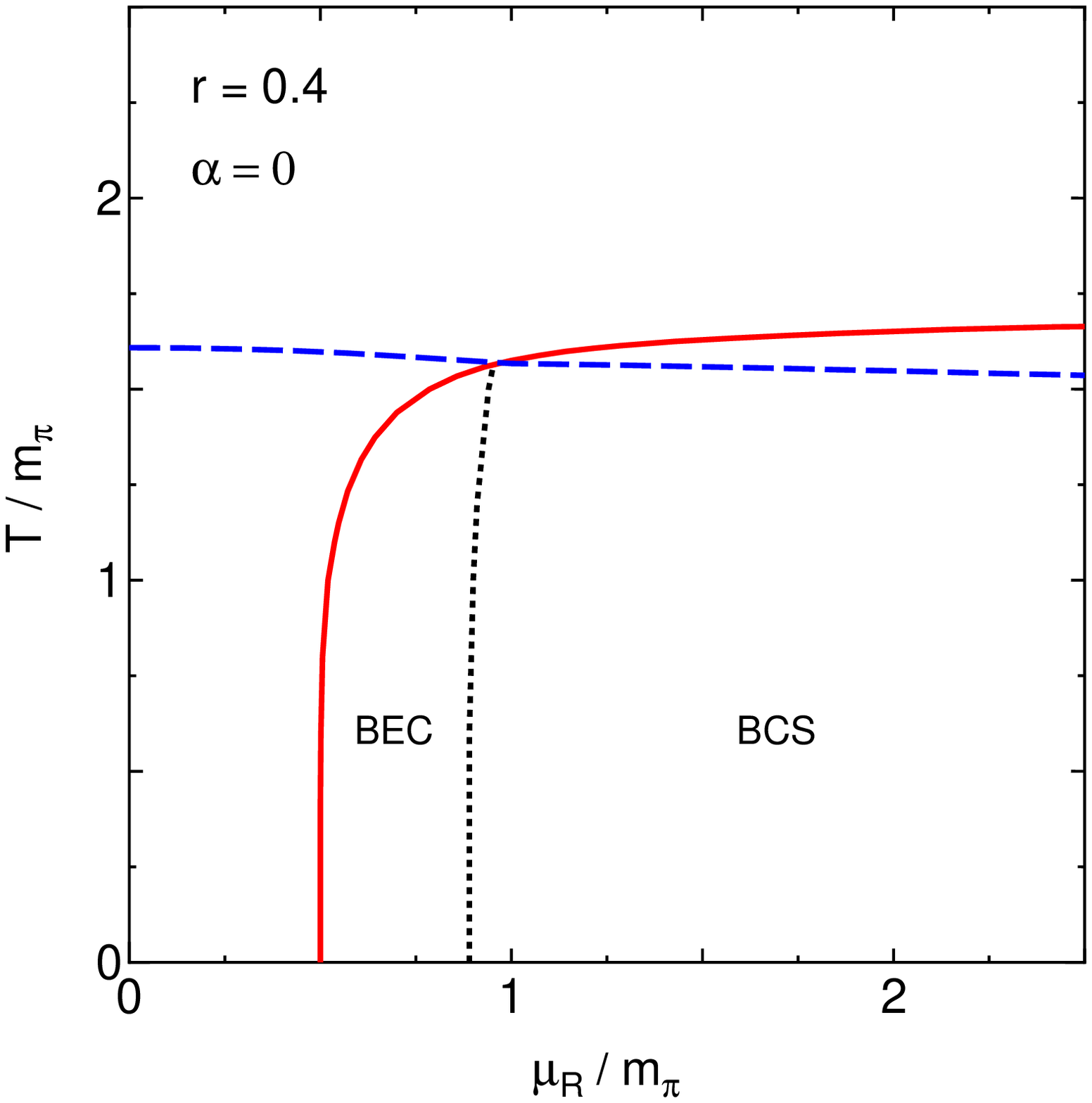}
\includegraphics[width=0.23\textwidth]{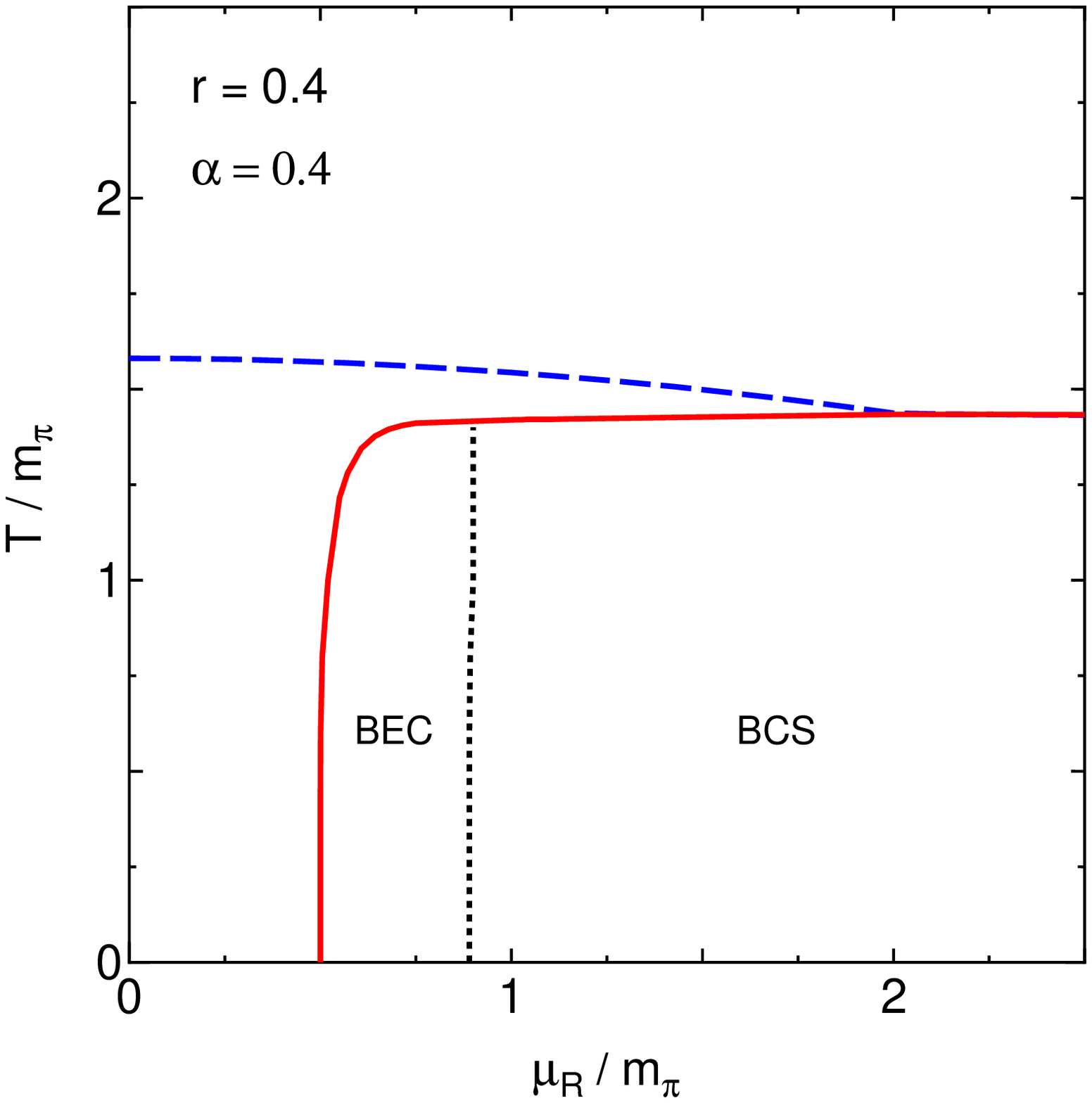}
\end{center}
\caption{
The phase diagram in the $\mu$-$T$ plane for
$\alpha=0$ (left panel) and $0.4$ (right panel).
Here the case of $r=0.4$ is taken.
The dotted, dashed and solid lines stand for the BCS-BEC crossover
($M=\mu$), the deconfinement crossover ($\Phi=0.5$) and the
superfluid/normal transition lines, respectively.
}
\label{PD-R}
\end{figure}

Finally we consider the phase diagram at real $\mu$.
Figure~\ref{PD-R} shows the BCS-BEC crossover ($M=\mu$), the
deconfinement crossover ($\Phi=0.5$) and the superfluid/normal
transition lines for two cases of $\alpha=0$ and $0.4$; see
Ref.~\cite{Brauner:2009} for the definitions of these transitions.
When $T$ is small, the superfluid/normal transition occurs at
$\mu=m_{\pi}/2$ as expected.
When $T > m_{\pi}$ and $\mu > m_{\pi}/2$, meanwhile, the
superfluid/normal transition takes place around $T=1.5 m_{\pi}$.
The transition line depends on
$\alpha$ rather strongly compared with the BCS-BEC crossover and the
deconfinement crossover line.
Thus the correlation between the chiral and deconfinement transitions is
important not only at $\mu^2=-(T\pi/2)^2$ but also at large real $\mu$
such as $\mu^2 > (m_{\pi}/2)^2$.

\section{Summary}
\label{Summary}

We have studied properties of two-color QCD at imaginary $\mu=i\theta T$
from the viewpoint of the RW periodicity, the pseudo-reality and
${\cal C}{\mathbb Z}_2$ symmetry.
Two-color QCD has ${\cal C}{\mathbb Z}_{2}$ symmetry at $\theta=\pm \pi/2$.
The PNJL model has the same properties as two-color QCD for the RW
periodicity, the pseudo-reality and ${\cal C}{\mathbb Z}_2$ symmetry.
The PNJL model is thus a good model to investigate two-color QCD at
imaginary $\mu$ concretely.
We have then investigated the nontrivial correlation between the
deconfinement and chiral transitions at imaginary $\mu$ for the
two-flavor case.

At $\theta=\pi/2$ and $T \ge T^c_{\rm RW}$, i.e., on the RW
phase-transition line, the spontaneous breaking of ${\cal C}{\mathbb
Z}_{2}$ symmetry takes place.
The ${\cal C}{\mathbb Z}_{2}$ symmetry breaking at the RW endpoint
$(\theta,T)=(\pi/2,T^c_{\rm RW})$ is continuously connected to the
deconfinement transition at $0 \le \theta <\pi/2$.
Thus the crossover deconfinement transition at $\theta=0$ is a remnant
of the ${\cal C}{\mathbb Z}_{2}$ symmetry breaking at the RW endpoint.

The order of the ${\cal C}{\mathbb Z}_{2}$ symmetry breaking
at the RW endpoint is nontrivial.
It cannot be determined by ${\cal C}{\mathbb Z}_{2}$ symmetry and the
pseudo-reality.
The order depends on the strength of the entanglement parameter $\alpha$,
i.e., the strength of the correlation between chiral and ${\cal
C}{\mathbb Z}_2$ symmetry breakings.
The order is second-order for small $\alpha$, but becomes first-order
for large $\alpha$.
The second-order nature is originated in the Polyakov-loop effective
potential.
Meanwhile, the first-order nature comes from the fact that chiral
symmetry breaking becomes first-order as a consequence of the strong
entanglement.
The order of ${\cal C}{\mathbb Z}_{2}$ symmetry breaking at the RW
endpoint is thus sensitive to the strength of the correlation between
chiral and ${\cal C}{\mathbb Z}_2$ symmetry breakings.
Finally, we have investigated the impact of $\alpha$ on the phase
diagram at real $\mu$.
The diagram, particularly the superfluid transition, is rather sensitive
to $\alpha$.
The determination of $\alpha$ is thus important for both real and
imaginary $\mu$.

At the present stage, we do not know how large $\alpha$ is, but it is
possible to determine the value of $\alpha$ from LQCD simulations at
$\theta=\pi/2$, particularly by seeing the correlation between chiral
and ${\cal C}{\mathbb Z}_2$ symmetry breakings.
It is interesting as a future work.
In two-color QCD, the correlation between the chiral and deconfinement
transitions is thus important for both real and imaginary $\mu$.
This is true also for three-color QCD.
This strongly suggests that understanding of three-color QCD at imaginary
$\mu$ is important to determine the phase diagram at real $\mu$.

\noindent
\begin{acknowledgments}
K.K. is supported by RIKEN Special Postdoctoral Researchers Program.
T.S. is supported by JSPS KAKENHI Grant Number 23-2790.
\end{acknowledgments}

\bibliography{Two-color.bib}

\end{document}